\documentclass[aps,pra,longbibliography,preprint,floatfix]{revtex4-1}

\usepackage[utf8]{inputenc}
\usepackage{amsmath, amssymb,graphicx}
\usepackage[cdot,mediumqspace,squaren]{SIunits}
\usepackage{hyperref}
\usepackage{xcolor}

\renewcommand{\vec}{\mathbf} 
\newcommand{\e}[1]{\text{e}^{#1}} 
\newcommand{\eps}{\varepsilon} 
\renewcommand{\k}{\vec{k}} 
\newcommand{\db}[1]{\overline{#1}} 

\graphicspath{{./images/},{./},{../images/}}
\begin{document}

\title{Chiral-Gain Photonics}
\author{Sylvain Lanneb\`{e}re\textsuperscript{1}}
\author{David E. Fernandes\textsuperscript{2}}
\author{Tiago A. Morgado\textsuperscript{1}}
\author{M\'{a}rio G. Silveirinha\textsuperscript{2}}
\email{To whom correspondence should be addressed:
mario.silveirinha@tecnico.ulisboa.pt}
 \affiliation{\textsuperscript{1}
Instituto de Telecomunica\c{c}\~{o}es and Department of Electrical Engineering, University of Coimbra, 3030-290 Coimbra, Portugal}
\affiliation{\textsuperscript{2}University of Lisbon -- Instituto
Superior T\'ecnico and
Instituto de Telecomunica\c{c}\~{o}es, 1049-001
Lisboa, Portugal}

\begin{abstract}
Here, we present an exploratory study of the potential applications of electrically biased materials that possess a nonreciprocal and non-Hermitian electromagnetic response analogous to the electronic response of field-effect transistors.
The most distinctive feature of such materials is their chiral-gain, meaning that their response can be active or dissipative depending on the handedness of the wave polarization. Here, we show how the chiral-gain can be harnessed  to develop novel electromagnetic devices with unique properties such as chiral lasers,  polarization-dependent mirrors, and coherent-perfect-absorber lasers. 
Furthermore, it is demonstrated that materials with chiral-gain can bypass a reciprocity constraint that typically limits the external coupling strength, thus facilitating the excitation of cavities with extremely large quality factors. 
\end{abstract}

\maketitle

\section{Introduction}
Photonic integrated circuits have the potential to revolutionize information processing devices by enhancing their speed, bandwidth, and power efficiency \cite{chrostowski_silicon_2015,komljenovic_heterogeneous_2016,karabchevsky_chip_2020}. However, their development has been slowed down by several challenges, particularly the absence of integrated light sources and amplifiers suitable for large scale production \cite{zhou_chip_2015,wang_novel_2017,zhou_prospects_2023,azzam_ten_2020}, as well as the lack of efficient integrated nonreciprocal components \cite{levy_chip_2002,caloz_electromagnetic_2018,asadchy_tutorial_2020}.

In a recent article \cite{lannebere_nonreciprocal_2022}, we introduced a nonlinear (meta-)material that mimics the response of semiconductor transistors. Such an idealized platform, which we designated by MOSFET-metamaterial, is characterized by a linearized optical response that is both nonreciprocal and non-Hermitian. These effects arise from the the interplay between material nonlinearities with a static electric bias. The most remarkable property of the metamaterial is that its non-Hermitian response is controlled by an \emph{indefinite} matrix (see below). Consequently, the interactions of a wave with the material can be either dissipative or exhibit gain, depending on the field-polarization.
In a series of subsequent articles, we demonstrated that transistor-like electromagnetic responses may be implemented in different ways, using transmission lines coupled with FETs \cite{fernandes_exceptional_2024}, in 2D engineered materials (strained twisted bilayer graphene) \cite{rappoport_engineering_2023}, and in 3D  low symmetry natural materials with a large Berry curvature dipole \cite{morgado_nonhermitian_2024}. Remarkably, the gain response of such platforms has chiral-properties, such that the material exhibits gain for a particular handedness of the optical field, while it is dissipative for the opposite polarization handedness \cite{lannebere_nonreciprocal_2022, rappoport_engineering_2023, hakimi_chiral_2023, morgado_nonhermitian_2024}.

The objective of this article is to present an exploratory study of the potential applications of such ``chiral-gain'' materials in the realm of nanophotonics.
Here, similar to \cite{lannebere_nonreciprocal_2022}, we restrict our attention to the case of an idealized  nonlinear material characterized by the constitutive relation  $ \vec{P} =\eps_0
\db{\chi}(\vec{E}) \cdot \vec{E}$ with a nonlinear susceptibility of the form
\begin{align}  \label{E:non_linear_susceptibility}
 \db{\chi}(\vec{E})&= \begin{pmatrix} \chi_{xx}(E_z) & 0 & 0 \\ 0 & \chi_{yy} & 0 \\ 0 & 0 & \chi_{zz}\end{pmatrix}
\end{align}
where $\vec{E}$ is the electric field and $\vec{P}$ is the polarization vector. As shown in \cite{lannebere_nonreciprocal_2022}, under a static electric bias with a nonzero $x$ component, the linearization of $\vec{P}$ leads to a (linearized) permittivity tensor of the form
\begin{align}  \label{E:permittivity_MOSFET_MTM}
\db{\eps} =\begin{pmatrix} \eps_{xx}  & 0 & \eps_{xz}  \\ 0 & \eps_{yy}  & 0 \\ 0 & 0 & \eps_{zz}\end{pmatrix}.
\end{align}
Remarkably, this permittivity tensor describes a nonreciprocal ($\db{\eps}^T\neq\db{\eps}$) \cite{altman_reciprocity_2011} and non-Hermitian ($\db{\eps}^\dagger\neq\db{\eps}$) material response.
Additionally, in the special case where the permittivity tensor is real-valued and frequency independent, the linearized response is also time-reversal ($\mathcal{T}$) invariant ($\db{\eps}^\ast=\db{\eps}$).
Here, $T$, $\ast$ and $\dagger$  represent  the transpose,  the complex conjugate and the Hermitian-conjugate operations, respectively.

The non-Hermitian response of the material is ruled by the matrix $\db{\eps}'' \equiv \frac{\db{\eps}   - \db{\eps}^\dag }{2i}$ \cite{lannebere_nonreciprocal_2022}. For the model of Eq. \eqref{E:permittivity_MOSFET_MTM}, supposing for simplicity that the diagonal elements are real-valued, the matrix $\db{\eps}''$ is given by:
\begin{align}  \label{E:permittivity_NH}
\db{\eps}'' &= \frac{1}{2i} \begin{pmatrix} 0 & 0 & \eps_{xz}\\ 0&0&0\\  -\eps_{xz}^\ast &0&0
\end{pmatrix}.
\end{align}
As the matrix trace vanishes, the matrix is indefinite. This means that some of its eigenvalues are positive, corresponding to a dissipative response, whereas other eigenvalues are negative corresponding to a gain response. It was shown in \cite{lannebere_nonreciprocal_2022} that when $\eps_{xz}$ is real-valued the relevant eigenvectors of the matrix $\db{\eps}''$ are $\hat{\vec{x}} \pm i\hat{\vec{z}}$. Thus, the polarization states associated with the dissipative/gain responses are circular and have opposite handedness. The polarization that activates the gain (dissipation) is controlled by the orientation of the electric field static bias. 

In our previous work, we showed that a material with chiral-gain can be used to combine optical isolation with amplification within the same device \cite{lannebere_nonreciprocal_2022}.
Here, we analyze how the chiral-gain can be harnessed for several potential nanophotonic applications.
First, in section \ref{sec:PT-symmetry} we characterize the spectrum of generalized indefinite-gain materials and discuss the conditions for their stability and the role of parity-time ($\mathcal{PT}$) symmetry.
In section \ref{sec:laser}, we investigate the application of chiral-gain materials in lasing. In section \ref{sec:reflection_mode}, we theoretically demonstrate that chiral-gain materials can be used as polarization-dependent amplifying mirrors. Next, in section \ref{sec:open_cavity}, it is shown that open cavities formed by chiral-gain media may be engineered to exhibit an arbitrarily high quality factor. 
Finally, in section \ref{sec:CPA_laser}, we demonstrate that such cavities are ideal for light-storage and that their operation is not limited by the reciprocity constraint.
This distinctive behavior is traced back to the unique analytical structure of the scattering matrix, which simultaneously accommodates a zero and a pole at the same frequency. This property delineates a regime where coherent perfect absorption (CPA or anti-lasing) \cite{chong_coherent_2010,wan_time-reversed_2011} and ``quasi-lasing'' are seamlessly integrated within the same resonator, similar to the behavior of ``CPA-lasers'' \cite{longhi_mathcalpt-symmetric_2010,chong_PT-symmetry_2011,longhi_pt-symmetric_2014,wong_lasing_2016} and their nonlinear counterparts \cite{longhi_time-reversed_2011,schackert_three-wave_2013}. 
We find that, under optimal conditions, the energy retention within the cavity is primarily influenced by the excitation duration rather than by the incoming wave amplitude.


\section{Bulk stability and $\mathcal{PT}$-symmetry}\label{sec:PT-symmetry}

Indefinite-gain materials are characterized by a matrix $\db{\eps}'' = \frac{\db{\eps}   - \db{\eps}^\dag }{2i}$ that is indefinite, meaning that the interactions between the material and a wave can be either dissipative or exhibit gain. 
Next, we study the stability of a bulk material described by the relative permittivity tensor 
\begin{align}   \label{E:permittivity_generalized_MOSFET_MTM}
\db{\eps} =\begin{pmatrix} \eps_{xx}  & 0 & \eps_{xz}  \\ 0 & \eps_{yy}  & 0 \\ \eps_{zx} & 0 & \eps_{zz}\end{pmatrix}.
\end{align}
This corresponds to a generalization of the MOSFET-metamaterial model introduced in \cite{lannebere_nonreciprocal_2022}. 
When $\eps_{xz}\neq \eps_{zx}$ ($\eps_{xz}\neq \eps_{zx}^\ast$) the response is nonreciprocal (non-Hermitian). Furthermore, when the diagonal-elements are real-valued (as assumed below) the matrix $\db{\eps}''$ is indefinite. For simplicity, in this article we neglect the frequency dispersion.

The natural modes of an indefinite-gain medium can be obtained by solving Maxwell equations in the frequency domain
 \begin{subequations}\label{E:Maxwell_equations}
\begin{align}
\nabla\times \vec{E}&= i\omega \mu_0 \vec{H} \\
\nabla\times \vec{H}&= -i\omega  \eps_0 \db{\eps} \cdot \vec{E} 
\end{align}
 \end{subequations}
where $\omega$ is the oscillation frequency. A time variation of the type $\e{-i\omega t}$ is implicit.
For continuous material invariant under arbitrary space translations, the bulk normal modes are plane waves with a spatial variation of the form $\e{i\k \cdot \vec{r}}$, where $\k$ is the wave vector and  $\vec{r}$ is the position vector.
To assess the material stability we characterize the plane waves with a real-valued wave vector and complex frequency  $\omega=\omega'+ i \omega''$. 
The real part of $\omega $ determines the time period $T=2\pi/\omega'$ of the natural oscillations and the imaginary part determines the amplitude variation with time. Since $\e{-i\omega t}=\e{-i\omega' t}\e{\omega'' t}$, it follows that $\omega''>0$ corresponds to a wave growing exponentially with time (unstable response) while $\omega''<0$ to a wave decaying with time (stable response).
 The plane wave solutions satisfy the wave equation $\k\times (\k\times \vec{E})+ \left(\omega^2/c^2\right) \db{\eps} \cdot \vec{E}=0$ where  $c$ is the speed of light in vacuum.

The non-Hermitian physics is ruled by the nontrivial eigenvectors of $\db{\eps}''$, which are  in the $xoz$ plane. Thereby, we focus our attention in the  propagation along the $y$ direction, as it maximizes the gain or loss interactions.
For $\k=k  \hat{\vec{y}}$,  there are two transverse modes characterized by 
\begin{subequations}\label{E:modes_bulk_generalized_MOSFET}
\begin{align}
\omega&=\omega_\pm \equiv \frac{kc}{\sqrt{\eps_\pm }}  \label{E:eigenmodes_bulk_generalized_MOSFET}\\
\vec{E}&\sim \vec{E}_\pm \equiv  \eps_{xz} \hat{\vec{x}} + \left(\eps_\pm-\eps_{xx}  \right)  \hat{\vec{z}}\label{E:eigenvectors_bulk_generalized_MOSFET}\\
\eps_\pm&=\frac{\eps_{xx}+\eps_{zz}\pm \sqrt{\gamma}}{2} 
\label{E:effective_permittivity_bulk_generalized_MOSFET}
\end{align}
\end{subequations}
where $\gamma=(\eps_{xx}-\eps_{zz})^2+4 \eps_{xz} \eps_{zx}$.

\begin{figure*}[!ht]
\centering
\includegraphics[width=0.5\linewidth]{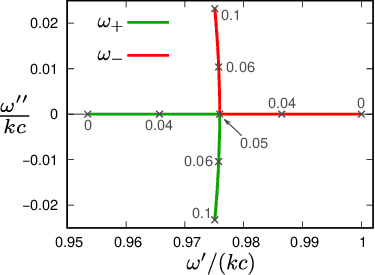}
     \caption{Locus in the complex plane of the normalized bulk natural frequencies $\omega_\pm/(k c)$ , when $\eps_{zx}$ varies in the range $0<\eps_{zx} < 0.1$. The insets indicate the value of $\eps_{zx}$ at the points marked with crosses. The remaining parameters used in the simulation are $\eps_{xx}=1$, $\eps_{zz}=1.1$ and $\eps_{xz}=-0.05$. The exceptional point ($\gamma=0$) is reached when $\eps_{zx}=-\eps_{xz}=0.05$.}
\label{fig:PT_illustration}
\end{figure*}

Consider first that the permittivity tensor \eqref{E:permittivity_generalized_MOSFET_MTM} is real-valued so that the response is invariant under a time-reversal. 
In this situation, $\gamma$ is also real-valued. 
When $\gamma>0$, both  $\eps_\pm$ [Eq. \eqref{E:effective_permittivity_bulk_generalized_MOSFET}]  and the natural frequencies $\omega_\pm$ [Eq.\eqref{E:eigenmodes_bulk_generalized_MOSFET}] are real-valued and non-degenerate. In this scenario $\omega''=0$ and the bulk material is stable. On the other hand, when $\gamma<0$, both $\eps_\pm$ and  $\omega_\pm$ become pairs of complex conjugates numbers, meaning that one mode is attenuated ($\omega''<0$) while the other is amplified ($\omega''>0$). In this case, which arises when $\eps_{xz}$ and $\eps_{zx}$ have opposite signs and their product satisfies $|\eps_{xz} \eps_{zx}|>\frac{1}{4}(\eps_{xx}-\eps_{zz})^2$, the bulk material is unstable. The transition from the stable to the unstable occurs when $\gamma=0$, which corresponds to an exceptional point where $\eps_\pm$, the eigenfrequencies [Eq. \eqref{E:eigenmodes_bulk_generalized_MOSFET}] and the eigenvectors [Eq. \eqref{E:eigenvectors_bulk_generalized_MOSFET}] coalesce and become degenerate. It is worth noting that when $\eps_{zz}=\eps_{xx}$, the analysis presented here reduces to the case discussed in \cite{buddhiraju_nonreciprocal_2020}. \\
The characteristics of the natural modes can be understood by noting that the material is invariant under a parity transformation $\mathcal{P}$ that flips the $y$ spatial coordinate $\mathcal{P} :(x,y,z)\to(x,-y,z)$. The permittivity tensor of a nonmagnetic and non-bianisotropic material invariant under $\mathcal{P}$ must satisfy $\db{\eps}(y)=\db{V}\cdot\db{\eps}(-y)\cdot\db{V}$, where $\db{V}=\hat{\vec{x}} \otimes \hat{\vec{x}} - \hat{\vec{y}} \otimes \hat{\vec{y}}+ \hat{\vec{z}} \otimes \hat{\vec{z}} $. This condition is satisfied by the permittivity tensor \eqref{E:permittivity_generalized_MOSFET_MTM}. Therefore, since the material is invariant under both  $\mathcal{P}$ and $\mathcal{T}$ transformations, it is also invariant under a $\mathcal{PT}$ transformation, the composition of the parity and time-reversal transformations. 
$\mathcal{PT}$ symmetric systems are known to possess a spectrum with mirror symmetry with respect to the real-frequency axis and to exhibit stable and unstable phases separated by an exceptional point \cite{bender_PTsymmetric_1999,bender_making_2007}.
When $\gamma>0$, the system is in the so-called unbroken $\mathcal{PT}$ symmetric phase, where all the eigenmodes are real-valued and the bulk material is stable. When $\gamma<0$, the system is in the broken $\mathcal{PT}$ symmetric phase, where the eigenmodes appear in complex conjugated pairs, corresponding to an unstable phase.\\
This behavior is illustrated in Fig. \ref{fig:PT_illustration} where the locus of the normalized natural frequencies $\omega_\pm/(k c)$ given by \eqref{E:eigenmodes_bulk_generalized_MOSFET} is represented in the complex plane as a function of $\eps_{zx}$. 
For the considered range of parameters, it is seen that when $0<\eps_{zx}<0.05$ the parameter $\gamma$ is positive and both frequencies $\omega_\pm$ lie on the real axis, implying that the system is stable and is in the unbroken $\mathcal{PT}$ symmetric phase. As $\eps_{zx}$ increases from 0 to 0.05, it can be observed that the eigenmodes gradually converge towards each other until they coalesce at the exceptional point associated with $\eps_{zx}=0.05$. 
For larger values of $\eps_{zx}$, the eigenmodes leave the real axis and become complex conjugates. In this case, the system enters the broken $\mathcal{PT}$ symmetric phase, where it becomes unstable, with waves growing exponentially in time until some nonlinear saturation mechanism limits the gain. \\

Both for the stable ($\gamma>0$) and unstable ($\gamma<0$) situations, the eigenvectors of the indefinite matrix $\db{\eps}''$ that controls the non-Hermitian interactions are  $\hat{\vec{x}} \pm i\hat{\vec{z}}$. Thus, when $\db{\eps}$ is real-valued, the wave polarizations that activate the gain/loss are always circularly polarized (chiral-gain). On the other hand, if either $\eps_{xz}$ or $\eps_{zx}$ are allowed to be complex-valued the eigenpolarizations that control the non-Hermitian interactions are in general elliptically polarized, corresponding to a more general form of indefinite-gain. For example, when $\eps_{zx}=0$ and $\eps_{xz}$ is pure-imaginary (as considered later in the article) the relevant eigenpolarizations are linearly polarized and are of the form $\hat{\vec{x}} \pm \hat{\vec{z}}$. \\  

In the rest of the article, we focus on the permittivity tensor given by \eqref{E:permittivity_MOSFET_MTM} with $\eps_{zx}=0$.
From the previous stability analysis, it is evident that such a response is always stable ($\gamma \geq 0$), even when $\eps_{xz}$ is allowed to be complex-valued. Furthermore, the exceptional point ($\gamma=0$), where the eigenvectors of the two transverse natural modes coalesce into a single linear polarization, is reached when $\eps_{xx} = \eps_{zz}$.  In this regime the indefinite-gain medium behaves as an ideal travelling wave amplifier, as discussed in detail in \cite{fernandes_exceptional_2024}. 
Throughout the remainder of the paper, we avoid this scenario by assuming that $\eps_{xx} \neq \eps_{zz}$, so that the material response is ``bounded''.

\section{Lasing}\label{sec:laser}
In this section, we illustrate how the non-Hermitian response provided by chiral-gain media can be harnessed for the creation of an electromagnetic oscillator, i.e. a laser. A related study based on a stack of 2D materials was recently put forward in \cite{hakimi_chiral_2023}. Next, we study laser-configurations based on 3D materials.

A laser is an oscillator that supports exponentially growing natural modes ($\omega''>0$) before it reaches the saturation (nonlinear) regime. Here, we consider a one dimensional closed cavity consisting of a chiral-gain medium slab positioned in between two mirrors. We investigate the natural modes of such a cavity for wave propagation along the $y$ direction.\\

The complex eigenmodes of the cavity can be determined by solving Maxwell equations \eqref{E:Maxwell_equations}. For propagation along the $y$ direction, we have $\nabla= \partial/\partial y \,\hat{\vec{y}}$, and these equations can be written in a matrix form as
\begin{align} \label{E:Maxwell_Schrodinger_equation}
i \frac{\partial}{\partial y}  \vec{f}(y)=\db{\vec{M}} \cdot \vec{f}(y)
\end{align}
where the four-component state vector  $\vec{f}$ is defined by
\begin{align} 
\vec{f}(y)= \begin{pmatrix}E_x(y) & E_z(y) & H_x(y) & H_z(y) \end{pmatrix}^T
\end{align}
with $E_x$, $E_z$, $H_x$, $H_z$ the $x$ and $z$ components of the electric and magnetic fields at position $y$. The $4\times4$ matrix $\db{\vec{M}}$ is given by
\begin{align}
\db{\vec{M}}=\begin{pmatrix}
0 & 0 & 0&  \omega \mu_0\\
0 & 0 & -\omega \mu_0 & 0\\
0 & -\omega  \eps_0 \eps_{zz} & 0 & 0\\
\omega  \eps_0  \eps_{xx}  & \omega  \eps_0  \eps_{xz}   & 0  & 0
           \end{pmatrix}.
\end{align}
Similar to a Schr\"odinger equation, the solution of Eq. \eqref{E:Maxwell_Schrodinger_equation}  is
\begin{align}\label{E:solutions_f}
\vec{f}(y)= \e{-i y \db{\vec{M}} }\cdot \vec{f}(0),
\end{align}
where the exponential $\e{-iy\db{\vec{M}}}$ is a $4\times4$ matrix that plays the role of a transfer matrix. The matrix exponential can be evaluated analytically (not shown here), or alternatively it can be numerically evaluated. Thus, the fields at a generic position $y$ within the material depend only on the fields calculated at $y=0$. \\

\subsection{Cavity closed by two PECs}
\begin{figure*}[!ht]
\centering
\includegraphics[width=0.3\linewidth]{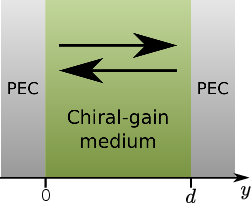}
     \caption{ Indefinite-gain medium cavity closed with two PEC walls at $y=0$ and $y=d$.}
\label{fig:MOSFET_cavity_PEC_PEC}
\end{figure*}

To begin, we consider the scenario where the two reflectors are perfect electric conducting (PEC) walls as represented in Fig. \ref{fig:MOSFET_cavity_PEC_PEC}.
In this case, the fields at each boundary of the cavity satisfy
\begin{subequations}
\begin{align}
\vec{f}(0)&=  H_x(0)\hat{\vec{u}}_3+ H_z(0) \hat{\vec{u}}_4 \label{E:boundary_condition_PEC_y=0}\\
\vec{f}(d)&= H_x(d)\hat{\vec{u}}_3+ H_z(d) \hat{\vec{u}}_4
\end{align}
\end{subequations}
where $\hat{\vec{u}}_i=(\delta_{i1}\;\delta_{i2}\;\delta_{i3}\;\delta_{i4})^T$ with $\delta_{ij}$ the Kronecker delta symbol, is a vector with the entry "1" at the $i$-th column. Taking into account that $\vec{f}(d)= \e{-id\db{\vec{M}}}\cdot \vec{f}(0)$ [Eq.\eqref{E:solutions_f}] and that $\vec{f}(d)\cdot \hat{\vec{u}}_1=\vec{f}(d)\cdot \hat{\vec{u}}_2=0$, we find after straightforward manipulations the following system of equations:
\begin{align}\label{E:characteristic_eq_PEC}
\begin{pmatrix}\hat{\vec{u}}_1\cdot\e{-id\db{\vec{M}}}\cdot \hat{\vec{u}}_3 & &  \hat{\vec{u}}_1\cdot\e{-id\db{\vec{M}}}\cdot \hat{\vec{u}}_4 \\
\hat{\vec{u}}_2 \cdot \e{-id\db{\vec{M}}}\cdot \hat{\vec{u}}_3& & \hat{\vec{u}}_2 \cdot \e{-id\db{\vec{M}}}\cdot \hat{\vec{u}}_4\end{pmatrix} \cdot \begin{pmatrix}H_x(0)\\H_z(0)\end{pmatrix}=\begin{pmatrix}0 \\ 0 \end{pmatrix}
\end{align}
\\
The dispersion equation is found by setting the determinant of the matrix in Eq. \eqref{E:characteristic_eq_PEC} equal to zero.
After simplifications, we obtain the following explicit result 
\begin{align}
  \sin \left( \frac{\omega}{c} d     \sqrt{ \eps_{xx}}\right) \sin \left(  \frac{\omega}{c}  d  \sqrt{ \eps_{zz}}\right)  = 0.
\end{align}
Clearly, the natural frequencies of the cavity of Fig. \ref{fig:MOSFET_cavity_PEC_PEC} are $\omega=\omega_n^x$ and $\omega=\omega_n^z$ with
\begin{subequations}
\begin{align}
  \omega_n^x&=  \frac{n \pi c}{d     \sqrt{ \eps_{xx}}} \label{E:resonant_frq_PEC_x}\\
  \omega_n^z&=  \frac{n \pi c}{d     \sqrt{ \eps_{zz}}} \label{E:resonant_frq_PEC_z}
\end{align}
\end{subequations}
and $n$ is an integer. Curiously, even though the bulk material can provide optical gain \cite{lannebere_nonreciprocal_2022}, the cavity eigenmodes are real-valued and consequently it is not possible to generate lasing relying simply on PEC walls.  To understand why, let us examine the time averaged net power exchanged between the wave and the material. For time-harmonic fields $\vec{E}(t) =\frac{1}{2}\left(\vec{E}_\omega \e{-i\omega t} + \vec{E}_\omega^\ast \e{i\omega t}\right)$ it is given by $\left\langle Q\right\rangle = \frac{\omega\eps_0}{2} \int  \vec{E}_\omega^\ast \cdot  \db{\eps}'' \cdot \vec{E}_\omega~dV$ (see the supplemental material of \cite{lannebere_nonreciprocal_2022}). For a chiral-gain medium described by the permittivity tensor  \eqref{E:permittivity_MOSFET_MTM} with  purely real diagonal elements, one has
\begin{align}  \label{E:power_transferred}
\left\langle Q\right\rangle  &= \frac{\omega\eps_0}{2} \int \mathrm{Im}\left\{ \eps_{xz} E_{\omega,x}^\ast  E_{\omega,z}\right\} ~dV.
\end{align}
From Eqs. \eqref{E:solutions_f}, \eqref{E:boundary_condition_PEC_y=0} and \eqref{E:characteristic_eq_PEC} the eigenvectors $\vec{E}_n^x$ and $\vec{E}_n^z$ associated with the natural modes $\omega_n^x$ and $\omega_n^z$, respectively, are:
\begin{subequations}
\begin{align}
  \vec{E}_n^x&= A^x \sin\left( \frac{n \pi y}{d}\right) \hat{\vec{x}} \\
  \vec{E}_n^z&= A^z \sin\left( \frac{n \pi y}{d}\right) \left(   \hat{\vec{x}} + \frac{\eps_{zz}-\eps_{xx}}{\eps_{xz}} \hat{\vec{z}} \right)
\end{align}
\end{subequations}
where $A^x$ and $A^z$ are normalization constants. As seen, the eigenmodes are linearly polarized
everywhere in the cavity when $\eps_{xz}$ is real-valued. 
When $\eps_{xz}$ is complex-valued, the eigenmode associated with $\omega_n^z$ becomes elliptically polarized. 
In either case, by substituting the explicit formulas for the fields into Eq. \eqref{E:power_transferred} ($\vec{E}_\omega\to\vec{E}_n^i$) one finds that $\left\langle Q\right\rangle=0$ for arbitrary $\eps_{xz}$. Thus, in this configuration the polarization of the eigenmodes is unable to unlock the gain and consequently there is no net transfer of energy from the material to the wave.\\
It is interesting to point out that the PEC mirrors preserve the $\mathcal{PT}$-symmetry of the bulk material, and thereby for a real-valued $\db{\eps}$  the spectrum must remain real-valued for an unbroken $\mathcal{PT}$-phase.

\subsection{Cavity closed by a PEC and a soft-hard boundary} \label{sec:kildal_closed_cavity}
\begin{figure*}[!ht]
\centering
\includegraphics[width=0.89\linewidth]{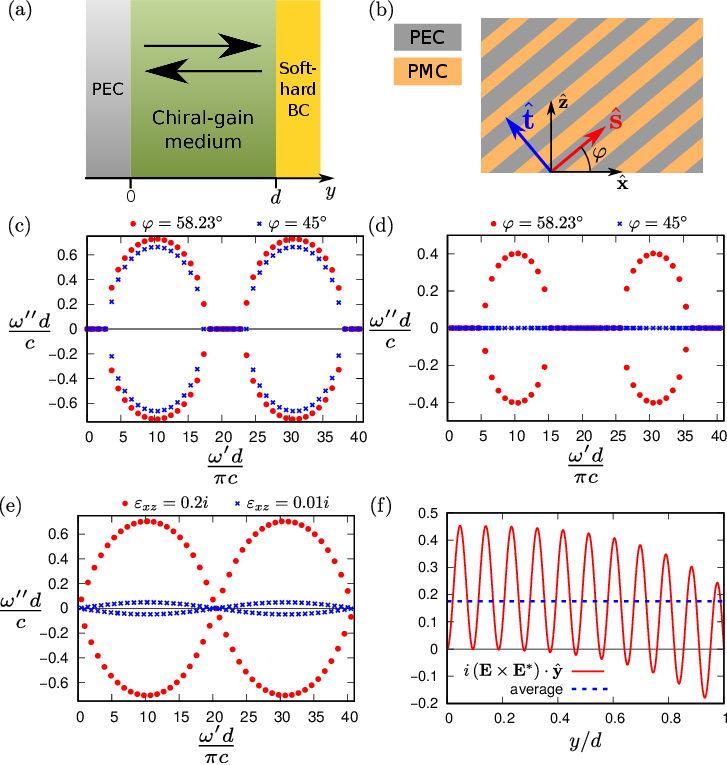}
     \caption{$\bf{(a)}$ Indefinite-gain medium cavity closed with a PEC wall at $y=0$ and a soft-hard boundary condition at $y=d$. $\bf{(b)}$ Sketch of a soft-hard boundary  implemented by alternating thin strips of PEC and PMC materials oriented along $\hat{\vec{s}}$. $\bf{(c)}$--$\bf{(e)}$ Locus  of the eigenfrequencies $\omega=\omega'+ i \omega''$ in the complex-plane for the cavity represented in (a).  The cross permittivity component is $\eps_{xz}=0.2$ in (c) and $\eps_{xz}=0.1$ in (d). The red dots (blue crosses) are calculated for an orientation of the soft-hard mirror $\varphi=58.23\degree$ ($\varphi=45\degree$). In (e) $\varphi=45\degree$ while $\eps_{xz}=0.2i$ (red dots) or $\eps_{xz}=0.01i$ (blue crosses). $\bf{(f)}$ $y$-component of the spin angular momentum (in arbitrary units) and its spatial average for the mode with the highest value of $\omega''$ for the blue crosses in (c). In all plots $\eps_{xx}=1$ and $\eps_{zz}=1.1$.}
\label{fig:MOSFET_cavity}
\end{figure*}

To extract energy from the medium we need to engineer the polarization of the cavity eigenmodes so that it ``overlaps'' the eigenvector of $\db{\eps}''$ that triggers the gain response. This can be done by replacing one of the PEC walls by a soft-hard
boundary (or Kildal’s mirror) \cite{kildal_artificially_1990,kildal_em_2003} as represented in Fig. \ref{fig:MOSFET_cavity} (a).
An ideal soft-hard boundary may be visualized as a set of alternating strips made of PEC and perfect magnetic conductor (PMC) [Fig. \ref{fig:MOSFET_cavity} (b)]. In practice, this type of boundary condition can be implemented using a corrugated plate mirror \cite{kildal_artificially_1990,kildal_em_2003}. Such mirror imposes that the electric and magnetic fields at the boundary satisfy
\begin{align}\label{E:boundary_condition_kildal_1}
\vec{E}\cdot\hat{\vec{s}} = \vec{H}\cdot\hat{\vec{s}}=0
\end{align}
where $\hat{\vec{s}}=\cos(\varphi)\hat{\vec{x}}+\sin(\varphi)\hat{\vec{z}}$ is a unit vector parallel to the direction of the strips as shown in Fig. \ref{fig:MOSFET_cavity} (b). The vector $\hat{\vec{s}}$ makes an angle $\varphi$ with the $x$ axis. It is useful to introduce the unit vector $\hat{\vec{t}}=-\sin(\varphi)\hat{\vec{x}}+\cos(\varphi)\hat{\vec{z}}$ that is orthogonal to $\hat{\vec{s}}$. Note that the combination of the PEC boundary with the Kildal boundary breaks the $\mathcal{P}$ and $\mathcal{PT}$ symmetries of the system. 

For the geometry of Fig. \ref{fig:MOSFET_cavity} (a), the PEC boundary condition imposes that the state vector  satisfies Eq. \eqref{E:boundary_condition_PEC_y=0} at $y=0$, whereas the soft-hard boundary condition at $y=d$ imposes that
\begin{align}
\vec{f}(d)\cdot \hat{\vec{s}}_E= \vec{f}(d)\cdot \hat{\vec{s}}_H=0
\end{align}
where $\hat{\vec{s}}_E=(s_x\; s_z \;0\;0)^T$ and $\hat{\vec{s}}_H=(0\;0\;s_x\; s_z)^T$ with $s_x=\hat{\vec{s}}\cdot \hat{\vec{x}}$ and $s_z=\hat{\vec{s}}\cdot \hat{\vec{z}}$. Using Eq. \eqref{E:solutions_f} it may be shown that the cavity fields are constrained as
\begin{align}
\begin{pmatrix}\hat{\vec{s}}_E\cdot\e{-id\db{\vec{M}}}\cdot \hat{\vec{u}}_3 & &  \hat{\vec{s}}_E\cdot\e{-id\db{\vec{M}}}\cdot \hat{\vec{u}}_4 \\
\hat{\vec{s}}_H \cdot \e{-id\db{\vec{M}}}\cdot \hat{\vec{u}}_3& & \hat{\vec{s}}_H \cdot \e{-id\db{\vec{M}}}\cdot \hat{\vec{u}}_4\end{pmatrix} \cdot \begin{pmatrix}H_x(0)\\H_z(0)\end{pmatrix}=\begin{pmatrix}0 \\ 0 \end{pmatrix}.
\end{align}
Again, the dispersion equation is found by setting the determinant of the matrix on the left-hand side equal to zero. After simplifications, we find that for $\eps_{xx}\neq\eps_{zz}$, the natural frequencies are given by the solutions of:  
\begin{align} 
   \beta_-\sin \left(\frac{\omega d }{c}   \left[\sqrt{\eps_{zz}}-\sqrt{\eps_{xx}}\right]\right)  -\beta_+ \sin \left(\frac{\omega d }{c}   \left[\sqrt{\eps_{zz}}+\sqrt{\eps_{xx}}\right]\right)   =0 \label{E:characteristic_eq_PEC_Kildal}
\end{align}
where
\begin{multline}
   \beta_\pm= \left[\left(\sqrt{\eps_{zz}} \pm \sqrt{\eps_{xx}} \right) \left[\sqrt{\eps_{zz}}\cos^2\left( \varphi \right) \pm \sqrt{\eps_{xx}}\sin^2\left( \varphi \right) \right]    - \eps_{xz} \cos\left( \varphi \right) \sin\left( \varphi \right) \right] \cdot\\ \left(\sqrt{\eps_{zz}} \mp \sqrt{\eps_{xx}} \right).
\end{multline}
For $\varphi=0$ or $\varphi=\pi/2$, i.e., when the strips are parallel to a coordinate axis, one has 
$\beta_+=\pm \beta_-$.
As the sum (or the difference) of two sine-functions can be written as the product of a sine function and a cosine function, it is clear that for $\varphi=0$ or $\varphi=\pi/2$ 
the natural frequencies of the cavity are purely real. For these specific orientations, the corrugated grid cannot induce lasing in the cavity. Interestingly, for other orientations of the corrugations, the eigenfrequencies may become complex-valued depending on the value of  $\eps_{xz}$.\\
We show in Fig. \ref{fig:MOSFET_cavity} (c)--(e) the solutions $\omega=\omega'+ i \omega''$ of Eq. \eqref{E:characteristic_eq_PEC_Kildal} in the complex plane (eigenfrequencies) for different values of the cross-coupling term $\eps_{xz}$ and various orientations of the corrugations $\varphi$.
For all the examined cases, the eigenfrequencies form a discrete (countable) set without any finite accumulation points in the complex plane.\\
For the reasons discussed above, the modal equation \eqref{E:characteristic_eq_PEC_Kildal} has real-valued solutions when $\beta_+=\pm \beta_-$.
Solving the equations $\beta_+=\pm \beta_-$ with respect to $\eps_{xz}$ one finds that $\eps_{xz}=\eps_{xz,1}$ or $\eps_{xz}=\eps_{xz,2}$ where $\eps_{xz,1}\equiv\left( \eps_{xx} - \eps_{zz}\right)\tan(\varphi)$ and $\eps_{xz,2}\equiv\left( \eps_{zz} - \eps_{xx}\right)\cot(\varphi)$. Our numerical analysis shows that these values determine exactly the gain thresholds required to have lasing. As $\eps_{xz,1}$ and $\eps_{xz,2}$ have opposite signs, they determine the threshold for positive and negative values of $\eps_{xz}$. Specifically, to have lasing one needs that $\eps_{xz}>\eps_{xz,\text{th}}^+$ or $\eps_{xz}<\eps_{xz,\text{th}}^-$ with 
\begin{subequations}
\begin{align}
\eps_{xz,\text{th}}^+\equiv \max\left\{\eps_{xz,1},\eps_{xz,2}\right\},\\
\eps_{xz,\text{th}}^-\equiv \min\left\{\eps_{xz,1},\eps_{xz,2}\right\}.
\end{align}
\end{subequations}
It can be checked that in these conditions $|\beta_+|<|\beta_-|$.
It is implicit that $\eps_{xx}\neq\eps_{zz}$, $\varphi\neq 0, \pi/2$.
Remarkably, for a fixed value of the permittivity, the gain threshold can be made as small as desired by selecting the appropriate angle $\varphi$. Moreover, as expected the thresholds are always different from zero, indicating that in the absence of cross-coupling the resonant frequencies remain real-valued.

For $\eps_{xz}$ real, the best orientation for the corrugations, i.e., the orientation that yields the largest amplification rate $\omega''$, depends on  the specific elements of the permittivity tensor.
For  small $\left| {{\varepsilon _{xz}}} \right|$, the optimal grid orientation is close to either $\varphi=0$ or $\varphi=\ 90 \degree$. On the other hand, for moderately large values of $\left| {{\varepsilon _{xz}}} \right|$ the optimal grid orientation approaches $\varphi = 45 \degree$, as illustrated in the following numerical example.  \\
Figures \ref{fig:MOSFET_cavity} (c)--(d) depict the cavity spectra calculated for positive cross-coupling with $\eps_{xz}=0.2$ and $\eps_{xz}=0.1$, respectively. We consider two grid orientations $\varphi=58.23\degree$ and $\varphi=45.0\degree$. The corresponding positive gain threshold values are  $\eps_{xz,\text{th}}^+\approx0.062$ for $\varphi=58.23\degree$ and $\eps_{xz,\text{th}}^+=0.1$ for $\varphi=45\degree$. It can be seen in Fig.  \ref{fig:MOSFET_cavity} (d) that when  $\eps_{xz}=\eps_{xz,\text{th}}^+$  all the  eigenfrequencies are real-valued (blue crosses). Conversely, for $\eps_{xz}>\eps_{xz,\text{th}}^+$,  some of the modes split into two branches with opposite imaginary part of $\omega$ [red dots in Figs. \ref{fig:MOSFET_cavity} (c) and (d) and blue crosses in (c)]. 
This pattern is repeated periodically in frequency with period $\Delta \omega_1 \equiv \pi c/ \left( d \left| \sqrt{\eps_{xx}}-\sqrt{\eps_{zz}}\right|\right)$ independent of $\varphi$.
For $\eps_{xz}=0.2$, the optimal $\varphi$ is close to the value $\varphi=58.23\degree$ used in Fig. \ref{fig:MOSFET_cavity} (c), whereas for $\eps_{xz}=0.1$ the optimal orientation is  close to $\varphi=67.5\degree$ (not shown). In Fig. \ref{fig:MOSFET_cavity} (f), we represent the wave spin angular momentum density (proportional to  the vector $i\vec{E}\times\vec{E}^\ast$ \cite{bliokh_extraordinary_2014}), as a function of the position in the cavity. We consider the mode with the highest $\omega''$ in the example of Fig. \ref{fig:MOSFET_cavity} (c) corresponding to the blue crosses. As seen, the spin angular momentum oscillates with the position in the cavity and takes both positive and negative values (associated with different handednesses). However, its average value aligns with the spin of the eigenmode of $\db{\eps}''$ that controls the gain (a specific circular polarization with handedness controlled by the sign of $\eps_{xz}$). We checked that this remains true for all the lasing modes (with $\omega''>0$).\\
When $\eps_{xz}$ is a pure-imaginary number, similar to the implementation proposed in \cite{fernandes_exceptional_2024}, $\beta_\pm$  are also complex-valued. 
In this situation, our numerical study shows that the cavity can always support solutions exponentially growing in time. Thus, when $\eps_{xz}$ is complex-valued, lasing is in principle possible as soon as the cross-coupling is different from zero as illustrated in Fig. \ref{fig:MOSFET_cavity} (e). The numerical simulations show that the best orientation for the corrugations is $\varphi=45\degree$.
The polarization of  lasing modes is now approximately linear because the eigenmode of $\db{\eps}''$ that controls the gain is of the type $\hat{\vec{x}} \pm \hat{\vec{z}}$.\\
It is interesting to note that when the modes move away from the real axis, the real-parts of adjacent eigenfrequencies are equally spaced by $\Delta \omega'=\Delta \omega_2$ with $\Delta \omega_2  \equiv 2\pi c/ \left( d \left[ \sqrt{\eps_{xx}}+\sqrt{\eps_{zz}}\right]\right)$ for modes in the same semi-plane (upper-half or lower-half frequency plane). 
For a real-valued $\eps_{xz}$, the spectrum has mirror symmetry about the horizontal axis [Figs. \ref{fig:MOSFET_cavity} (c)--(d)], analogous to $\mathcal{P} \mathcal{T}$ symmetric systems. We will explain this property in the next subsection with the help of image theory. In contrast, for a complex-valued $\eps_{xz}$, the eigenfrequencies in the different semi-planes are not related by complex-conjugation [Fig. \ref{fig:MOSFET_cavity} (e)].

\subsection{Cavity closed by two Kildal's mirrors}
\begin{figure*}[!ht]
\centering
\includegraphics[width=0.89\linewidth]{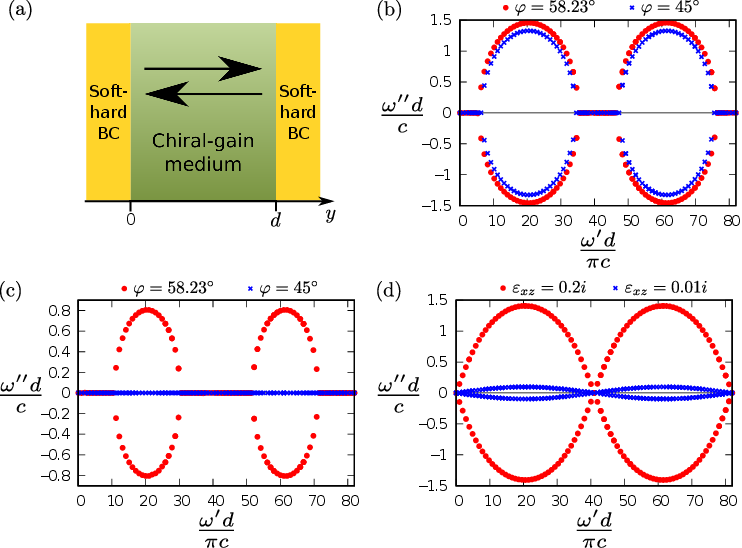}
     \caption{$\bf{(a)}$  Indefinite-gain medium cavity closed with two soft-hard mirrors at $y=0$ and $y=d$. $\bf{(b)}$--$\bf{(d)}$ Locus of the cavity eigenfrequencies   $\omega=\omega'+ i \omega''$ in the complex plane.  The cross permittivity is $\eps_{xz}=0.2$ in (b) and $\eps_{xz}=0.1$ in (c). The grid orientation is $\varphi=58.23\degree$ (red dots) or $\varphi=45\degree$ (blue crosses). The gain thresholds are the same as in Fig.\ref{fig:MOSFET_cavity}. In (d) $\varphi=45\degree$ and $\eps_{xz}=0.2i$ (red dots) or $\eps_{xz}=0.01i$ (blue crosses). In all plots $\eps_{xx}=1$ and $\eps_{zz}=1.1$.}
\label{fig:MOSFET_cavity_two_kildal}
\end{figure*}

Suppose now that the PEC wall at $y=0$ is replaced by a second soft-hard mirror as illustrated in Fig. \ref{fig:MOSFET_cavity_two_kildal} (a). For simplicity, it is assumed that the two soft-hard mirrors have the same orientation. When $\eps_{xz}$ is real-valued, the cavity is $\mathcal{PT}$-symmetric . 

Interestingly, the modes of this cavity can be split into $\bf{E}$-even and $\bf{E}$-odd modes, with the even (odd) solutions associated with a system where a PMC (PEC) mirror is placed at the center of the cavity. Thereby, the spectrum of this resonator is
the direct sum of the spectrum of a PEC/Kildal resonator with length $L/2$, already studied in subsection \ref{sec:kildal_closed_cavity}, and the spectrum of a PMC/Kildal resonator, also with length $L/2$. 
This observation explains why the spectrum of the PEC/Kildal cavity exhibits mirror symmetry when $\eps_{xz}$ is real-valued. Indeed, due to image theory these modes can be as well regarded as modes of a $\mathcal{PT}$-symmetric cavity with two Kildal mirrors. 

For the resonator with two Kildal mirrors,  the boundary conditions at $y=0$ imply that
\begin{align}
\vec{f}(0)=A_E \hat{\vec{t}}_E+A_H \hat{\vec{t}}_H
\end{align}
where $A_E$ and $A_H$ are some unknown coefficients, $\hat{\vec{t}}_E=(t_x\; t_z \;0\;0)^T$ and $\hat{\vec{t}}_H=(0\;0\;t_x\; t_z)^T$ with $t_x=\hat{\vec{t}}\cdot \hat{\vec{x}}$ and $t_z=\hat{\vec{t}}\cdot \hat{\vec{z}}$.
Imposing the soft-hard boundary conditions at $y=d$ one finds that:
\begin{align}\label{E:characteristic_eq_Kildal_Kildal}
\begin{pmatrix}\hat{\vec{s}}_E\cdot\e{-id\db{\vec{M}}}\cdot \hat{\vec{t}}_E & &  \hat{\vec{s}}_E\cdot\e{-id\db{\vec{M}}}\cdot \hat{\vec{t}}_H \\
\hat{\vec{s}}_H \cdot \e{-id\db{\vec{M}}}\cdot \hat{\vec{t}}_E & & \hat{\vec{s}}_H \cdot \e{-id\db{\vec{M}}}\cdot \hat{\vec{t}}_H \end{pmatrix} \cdot \begin{pmatrix}A_E\\A_H \end{pmatrix}=\begin{pmatrix}0 \\ 0 \end{pmatrix}.
\end{align}
The dispersion equation is obtained using the same procedure as before.\\
Consistent with image theory, our numerical studies  show that  the resonant frequencies for this configuration exhibit the same qualitative behavior as in the PEC/Kildal cavity  of Fig.\ref{fig:MOSFET_cavity} (a). In particular, we find that resonant frequencies with a positive imaginary part always exist for a pure imaginary $\eps_{xz}$ [see Fig. \ref{fig:MOSFET_cavity_two_kildal} (d)], and are only allowed  above some positive and negative thresholds $\eps_{xz,\text{th}}^\pm$  for a real-valued $\eps_{xz}$ [see Fig. \ref{fig:MOSFET_cavity_two_kildal} (c)].
The gain thresholds for lasing are coincident with the ones determined in the previous section.
The optimal orientation for the corrugations for a real-valued $\eps_{xz}$ is also unchanged as compared to the previous subsection.
The main difference as compared to the PEC/Kildal cavity is the peak gain rate of the lasing modes, which is two times larger with the two Kildal mirrors.
Furthermore, the frequency period of the spectrum is also two times larger than before ($2 \Delta \omega_1$). These properties are all a simple consequence of image theory. 

In summary, we demonstrated that optical cavities based on a indefinite-gain media terminated by a Kildal's mirror and a PEC or two Kildal's mirrors may be used to implement electrically pumped chiral lasers. The configuration with two Kildal mirrors is the most advantageous as it provides the largest amplification rate.


\section{Chiral-gain Mirrors} \label{sec:reflection_mode}
\begin{figure*}[!ht]
\centering
\includegraphics[width=0.8\linewidth]{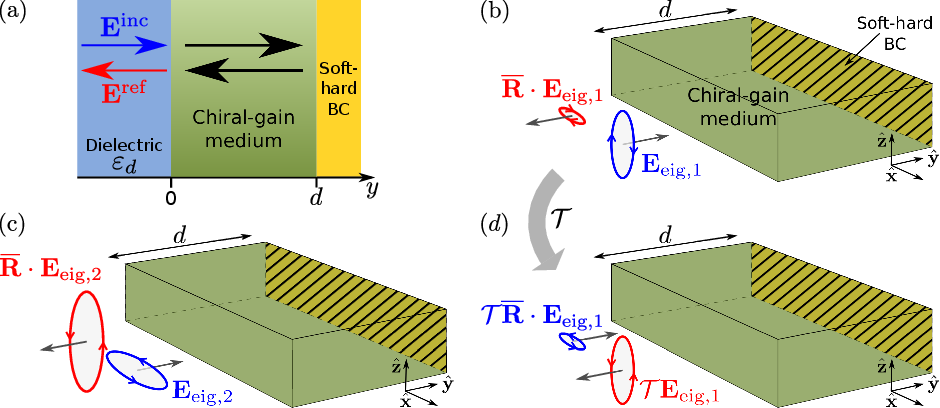}
     \caption{$\bf{(a)}$ A chiral-gain medium slab of thickness $d$ is terminated by a Kildal’s mirror. The material slab is illuminated along the normal direction by a plane wave $\vec{E}^\text{inc}$. $\bf{(b)}$-$\bf{(c)}$ Schematics depicting the working principle of the non-Hermitian mirror when $|\rho_1|^2<1$ (panel (b), corresponding to attenuation) and $|\rho_2|^2>1$ (panel (c), corresponding to gain). $\bf{(d)}$ Illustration of the mirror response when a time-reversal transformation ($\mathcal{T}$) is applied to  scenario (b). In all plots, the incident (reflected) wave is represented in blue (red).}
\label{fig:open_MOSFET_cavity}
\end{figure*}

In this section, we analyze another application of chiral-gain materials, namely the realization of non-Hermitian mirrors that provide either amplification or attenuation, depending on the wave polarization handedness. To this end, we consider the plane wave excitation of a chiral-gain medium slab backed by a mirror, as illustrated in Fig. \ref{fig:open_MOSFET_cavity} (a). We focus on the case where $\eps_{xz}$ is real-valued, so that the system is time-reversal symmetric.
The mirror is taken as a corrugated wall described by soft-hard boundary conditions, while the surrounding medium is a dielectric of permittivity $\eps_d$, for example air. As illustrated in Fig. \ref{fig:open_MOSFET_cavity} (b)--(c), the non-Hermitian mirror can be used to selectively amplify or attenuate the incoming wave, depending on its polarization state. This can be useful, for instance, to enhance the back-scattering of some object with a specific polarization signature.\\
It is assumed that the  plane wave propagates along the $+y$-direction and impinges on the slab along the normal direction. Let $ \vec{E}_t^\text{inc}=E_x^\text{inc} \hat{\vec{x}}  + E_z^\text{inc} \hat{\vec{z}}$ be the complex amplitude of the (transverse) incident field and $ \vec{E}_t^\text{ref}$ the complex amplitude of the (transverse) reflected field. At the $y=0$ interface  these fields are related by
 \begin{align}\label{E:E_reflected}
\vec{E}_t^\text{ref}(y=0)&=\db{\vec{R}}\cdot \vec{E}_t^\text{inc}(y=0)
\end{align}
where $\db{\vec{R}}$ is the 2$\times$2 reflection matrix, whose expression is derived in appendix \ref{sec:reflection_matrix} [Eq. \eqref{E:reflection_matrix}].
The power reflected by the chiral-gain  slab is proportional to 
\begin{align}\label{E:E_reflected_square}
|\vec{E}_t^\text{ref}|^2 &=\vec{E}_t^{\text{inc},\ast} \cdot \db{\vec{R}}^\dagger\cdot\db{\vec{R}} \cdot \vec{E}_t^\text{inc}
\end{align}
The reflectance matrix $\db{\mathcal{R}}=\db{\vec{R}}^\dagger\cdot\db{\vec{R}}$  plays a crucial role in our analysis. It is a positive definite Hermitian matrix, meaning that its eigenvectors $\vec{E}_{\text{eig},1}$ and $\vec{E}_{\text{eig},2}$ form an orthogonal basis, while the eigenvalues $|\rho_1|^2$ and $|\rho_2|^2$ are positive numbers.
It is assumed, without loss of generality, that $|\rho_1|^2\leq|\rho_2|^2$. Clearly, the reflectance $|\rho|^2\equiv|\vec{E}_t^\text{ref}|^2/|\vec{E}_t^\text{inc}|^2 $  is bounded by the eigenvalues of $\db{\mathcal{R}}$, namely $|\rho_1|^2\leq|\rho|^2\leq|\rho_2|^2$. Additionally, the  eigenvectors $\vec{E}_{\text{eig},1}$ and $\vec{E}_{\text{eig},2}$ determine the polarizations of the incident field that minimize and maximize the reflected signal power, respectively.
It is important to note that $ \vec{E}_{\text{eig},i}$ are not eigenvectors of the reflection matrix, which means that the reflected wave may have a polarization different from that of the incident wave when $\vec{E}_t^\text{inc}= \vec{E}_{\text{eig},i}$. \\
In a lossless and passive system (two-port microwave network), energy conservation imposes
that the power reflected by the network must be identical to the incident power. Thus, in the conservative case one has $|\rho_1|^2=|\rho_2|^2=1$. On the other hand, non-Hermitian systems are not constrained by energy conservation, and hence it is possible to have
$|\rho_i|^2>1$ ($|\rho_i|^2<1$), which corresponds to amplification (attenuation) of the incident wave.
\begin{figure*}[!ht]
\centering
\includegraphics[width=0.8\linewidth]{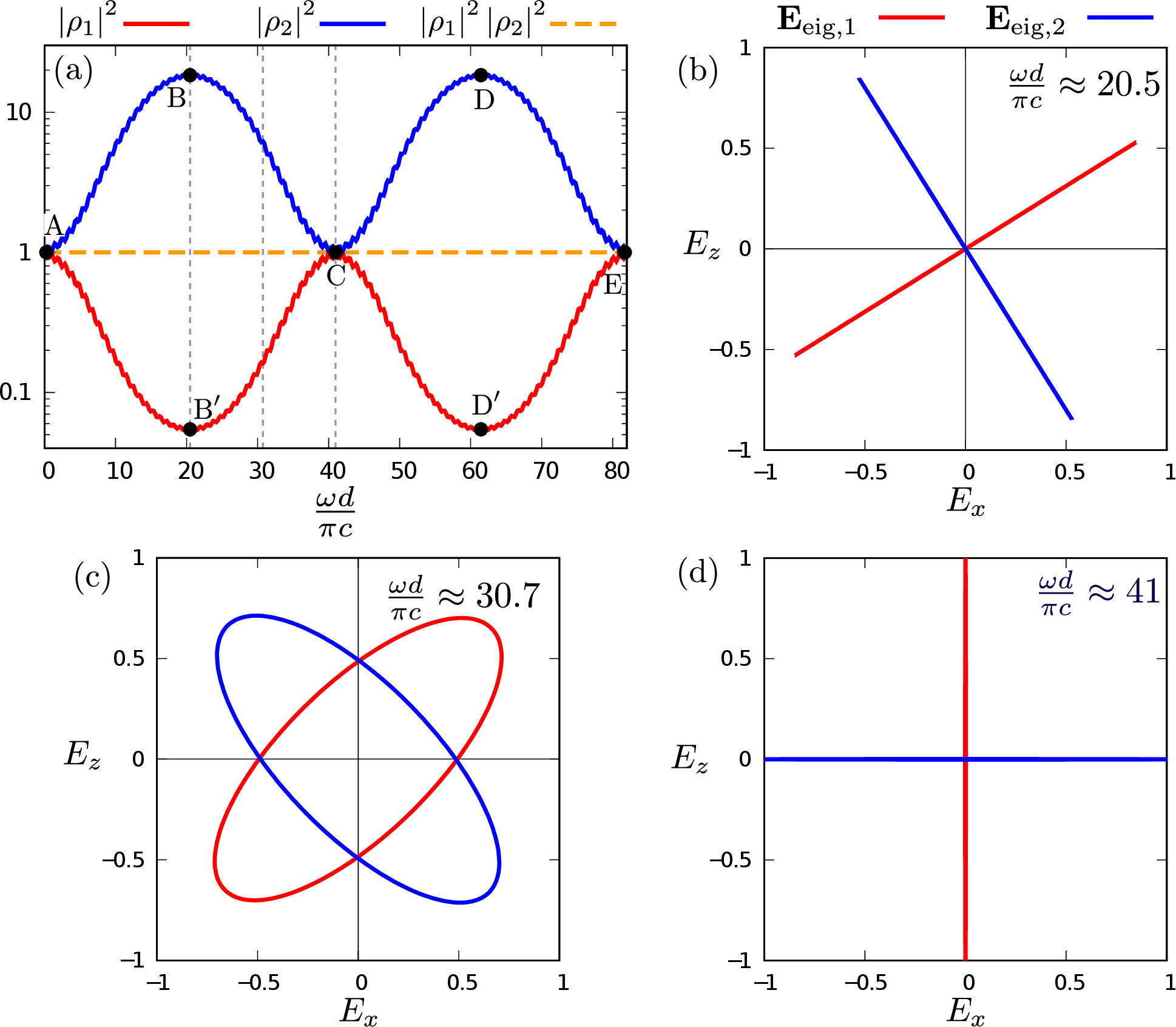}
       \caption{$\bf{(a)}$ Plot of the eigenvalues $|\rho_1|^2$,  $|\rho_2|^2$ of $\db{\mathcal{R}}$ and their product  $|\rho_1|^2 |\rho_2|^2$ (note the logarithmic scale), for $\eps_{d}=1$, $\eps_{xx}=1$, $\eps_{zz}=1.1$, $\eps_{xz}=0.1$, and $\varphi=45^\degree$. A, B, \dots, E represent the high-symmetry points within the plot. $\bf{(b)}$-$\bf{(d)}$ Polarization curves for the eigenvectors $ \vec{E}_{\text{eig},1}$, $ \vec{E}_{\text{eig},2}$ in the $xoz$ plane for three different frequencies  marked with vertical dashed lines in (a).}
\label{fig:power_reflected}
\end{figure*}
Figure \ref{fig:power_reflected} shows a plot of the eigenvalues and eigenvectors of $\db{\mathcal{R}}$ for a ($\mathcal{T}$-invariant) chiral-gain medium slab surrounded by air ($\eps_{d}=1$). As seen, the eigenvalues are different from unity and have values that oscillate periodically with the frequency or with the slab thickness. 
As illustrated in Fig. \ref{fig:open_MOSFET_cavity} (b)--(c), in a wide range of frequencies, polarization 1 (red lines) can be strongly absorbed by the slab whereas in the same frequency range (the orthogonal) polarization 2 (blue lines) is strongly amplified leading to a reflected wave with amplitude larger than the incoming wave. This behavior is possible as soon as $\eps_{xz}$ is different from zero (no threshold). Furthermore, it can be checked that the orientation for the Kildal's mirror corrugations that provides the strongest amplification is the same as the orientation that gives the largest $\omega''$ in the closed PEC/Kildal cavity. The maximum power gain is  18.7 for the example of Fig. \ref{fig:power_reflected} (corresponding to point B of the plot) with $\varphi=45^\degree$, whereas it can reach  37.5 when $\varphi=67.5\degree$ (not shown). 

Quite remarkably the product $|\rho_1|^2 |\rho_2|^2$ is equal to the unity meaning that when one of the eigenvectors is amplified the corresponding orthogonal polarization is attenuated in the same proportion. The property $|\rho_1|^2 |\rho_2|^2=1$ is a consequence of time-reversal symmetry. Indeed,  under a time-reversal the roles of the incident and reflected waves are swapped [see Fig. \ref{fig:open_MOSFET_cavity} (b) and (d)] \cite{silveirinha_time-reversal_2019,fernandes_role_2022}. Thereby, the maxima and the minima of the reflectance are necessarily interchanged under a time-reversal $|\rho|^2\to 1/|\rho|^2$, implying that $|\rho_1|^2 = 1/|\rho_2|^2$ . Note that this result demonstrates that  $ \det \db{\mathcal{R}} = 1 = \left| \det \db{\vec{R}}\right|$.

Furthermore, the previous argument also implies that if the field reflected by wave 1, $\db{\vec{R}}  \cdot {{\vec{E}}_{{\rm{eig,1}}}}$, (i..e, by the wave ${\vec{E}}_{{\rm{eig,1}}}$ that activates maximum attenuation) is sent back to the material slab then it it will generate maximum gain. 
Thereby, it follows that ${\left( \db{\vec{R}}  \cdot {\vec{E}_{\rm{eig,1}}} \right)^\ast}\sim {{\vec{E}}_{{\rm{eig,2}}}}$. 
A similar identity holds true with the indices 1 and 2 interchanged. We took into account that under a time-reversal the  electric field is conjugated. 
Thus, we have shown that the field reflected by one of the eigenvectors is proportional to the time-reversed version of the other eigenvector. Formally, this can be expressed as:
\begin{align} \label{E:reflected_eigenmode}
\db{\vec{R}}\cdot  \vec{E}_{\text{eig},j}= |\rho_j| \e{i \theta_j} \vec{E}_{\text{eig},l}^\ast, \qquad j\neq l
\end{align}
with $\theta_j$ some phase. It is implicit in this discussion that $\omega$ is real-valued and that the system is non-Hermitian ($|\rho_1|\ne |\rho_2|$). The eigenvectors amplitude is normalized to the unity.

The polarization curves of the eigenvectors $ \vec{E}_{\text{eig},1}$, $ \vec{E}_{\text{eig},2}$ are depicted in Fig. \ref{fig:power_reflected} (b)--(d) for different values of the normalized frequency $\omega d /(\pi c)$ . 
The incident fields that maximize the amplification/absorption are generally elliptically polarized [see Fig. \ref{fig:power_reflected} (c)], except at some special high-symmetry points. Due to the orthogonality of the eigenvectors, the handedness of the wave that activates the dissipation,  $ \vec{E}_{\text{eig},1}$, is always the opposite of the handedness of the wave that activates the gain, $ \vec{E}_{\text{eig},2}$. 
From Eq. \eqref{E:reflected_eigenmode}, the polarization curve of the field reflected by wave 1 is coincident (apart from a scale factor and the absolute sense of rotation) with the polarization curve of wave 2. In particular, the polarization curves of an incident wave (1 or 2) and of the corresponding reflected wave always differ by a 90$\degree$ rotation.

Counter-intuitively, the frequency for which the mirror amplification/attenuation is maximized [points B, B', D and D' in Fig. \ref{fig:power_reflected} (a)], corresponds to linearly polarized incident wave [Fig. \ref{fig:power_reflected} (b)], with the electric field forming an angle of approximately $30^\degree$ with respect to the coordinate axes. This is possible because the field polarization inside the material slab is controlled not only by the polarization of the incident wave, but also by the anisotropy of the material and by the anisotropy of the Kildal-mirror.   Consequently, there is no strict correlation between the polarization of the incident wave that triggers the maximum gain and the eigenvector of the matrix $\db{\eps}''$ that maximizes the gain response of the medium. Thereby, an incident linearly polarized wave can unlock the chiral-gain within the material. It is also important to note that any incident polarization can be decomposed into the basis vectors $\vec{E}_{\text{eig},1}$ and $\vec{E}_{\text{eig},2}$. In cases of very strong gain, the component associated with $\vec{E}_{\text{eig},2}$ typically dominates the scattered field, even if it is weak in the incident wave. Therefore, there is no universal relationship between the polarization handedness of the incident field and the type of non-Hermitian (dissipative or amplifying) mirror response.
 
While the configuration of Fig. \ref{fig:open_MOSFET_cavity} provides the strongest non-Hermitian effects, it is also feasible to achieve chiral-amplification when the Kildal's mirror is replaced by a PEC mirror. For the parameters used in Fig. \ref{fig:power_reflected}, the chiral-gain slab backed by a PEC mirror yields a maximum power gain of 6 (not shown), which is roughly six times smaller than the maximum amplification  achievable with the Kildal's mirror. The fact that the slab backed by the PEC wall can provide amplification while a cavity enclosed by two PEC mirrors [Fig.\ref{fig:MOSFET_cavity_PEC_PEC}] cannot produce lasing does not result in any contradiction. This difference of behavior can be attributed to the polarization of the field inside the material slab. In the closed cavity, the polarization is controlled only by the boundary conditions. On the other hand, in the open system, the polarization of the incident field becomes a free parameter that can be harnessed to achieve either amplification or absorption.\\
In summary, it was demonstrated that a chiral-gain medium  can be used to realize polarization-controlled non-Hermitian mirrors. The degree of amplification/absorption is highly dependent on the polarization and frequency of the incident wave as well as on the slab thickness.

\section{Modes of the open cavity} \label{sec:open_cavity} 
Next, we study the natural modes supported by an open cavity, i.e., by a system with the same
geometry as in section \ref{sec:reflection_mode}, without the incoming wave. The natural modes of passive open systems are characterized by complex-valued  frequencies with $\omega''<0$, corresponding to an exponential decay in time. The oscillation damping is caused either by material absorption or by  radiation loss \cite{silveirinha_trapping_2014, hsu_bound_2016}. Here, we want to study if it is possible to exploit the chiral-gain to compensate these effects and engineer open resonators with high-quality factors.\\
For an open-resonator, the field in the dielectric-region ($y<0$) is an outgoing plane wave. In particular, at the interface ($y=0^-$) with the chiral-gain medium the state vector is of the form:
\begin{align}
\vec{f}_\text{out}(0)
=\begin{pmatrix}   \vec{E}_\text{out} \\  -\frac{\sqrt{\eps_d}}{\eta_0}   \db{\vec{J}}  \cdot \vec{E}_\text{out} \end{pmatrix}
\end{align}
where $\vec{E}_\text{out}$ is the transverse outgoing electric field and $\db{\vec{J}}$ is a $2 \times 2$ matrix defined as in Eq. \eqref{E:matrix_J}. We used Eq. \eqref{E:H_transverse_dielectric} to obtain this result.

The dispersion of the natural modes can be found by matching $\vec{f}_\text{out}$, with the field in the chiral-gain slab ($y=0^+$) terminated with a Kildal’s mirror [Eq. \eqref{E:f_zero_kildal}].This procedure yields the following dispersion equation
\begin{align}
\det\left[\begin{pmatrix} \hat{\vec{u}}_3 \cdot \e{i d \db{\vec{M}} }\cdot \hat{\vec{t}}_E  & \hat{\vec{u}}_3 \cdot \e{i d \db{\vec{M}} }\cdot \hat{\vec{t}}_H \\\hat{\vec{u}}_4 \cdot \e{i d \db{\vec{M}} }\cdot \hat{\vec{t}}_E  & \hat{\vec{u}}_4 \cdot \e{i d \db{\vec{M}} }\cdot \hat{\vec{t}}_H \end{pmatrix} + \frac{\sqrt{\eps_d}}{\eta_0} \db{\vec{J}} \cdot \begin{pmatrix}
 \hat{\vec{u}}_1 \cdot \e{i d \db{\vec{M}} }\cdot \hat{\vec{t}}_E  & \hat{\vec{u}}_1 \cdot \e{i d \db{\vec{M}} }\cdot \hat{\vec{t}}_H \\ \hat{\vec{u}}_2 \cdot \e{i d \db{\vec{M}} }\cdot \hat{\vec{t}}_E  & \hat{\vec{u}}_2 \cdot \e{i d \db{\vec{M}} }\cdot \hat{\vec{t}}_H \end{pmatrix} \right]=0
\end{align}
The roots of this equation give the complex eigenfrequencies $\omega=\omega'+i\omega''$ of the open cavity. The qualitative behavior of these solutions with respect to the cross-coupling coefficient $\eps_{xz}$ is analogous to the behavior observed in the closed cavity of Sect. \ref{sec:kildal_closed_cavity}. The main difference between the two configurations can be understood by comparing the spectra obtained with $\eps_{xz}=0$. Evidently, a closed cavity  with  $\eps_{xz}=0$ has all  the eigenfrequencies on the real-frequency axis.
\begin{figure*}[!ht]
\centering
\includegraphics[width=0.9\linewidth]{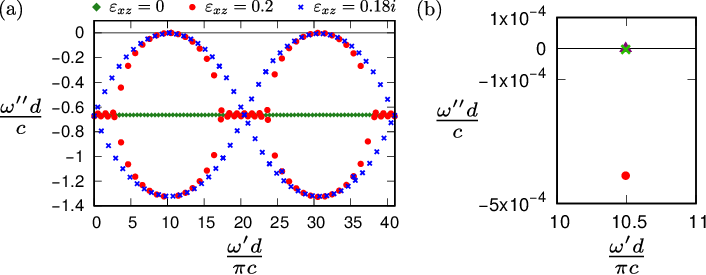}
       \caption{$\bf{(a)}$ Locus of the eigenfrequencies of the open cavity represented in Fig. \ref{fig:open_MOSFET_cavity} (a) for $\eps_{d}=3$, $\eps_{xx}=1$, $\eps_{zz}=1.1$, $\varphi=45^\degree$ and no cross-coupling $\eps_{xz}=0$ (green diamonds), a real-valued cross-coupling $\eps_{xz}=0.2$ (red dots) and a pure-imaginary cross-coupling $\eps_{xz}=0.18i$ (blue crosses). $\bf{(b)}$ Zoom in view of the spectrum near the real-frequency axis for  $\eps_{xz}=0.2$ and $\varphi=45^\degree$ (red dot), for $\eps_{xz}=0.2$ and $\varphi=45.0395^\degree$ (green cross) and for $\eps_{xz}=0.2001382$ and $\varphi=45^\degree$ (purple triangle).}
\label{fig:mode_open_cavity}
\end{figure*}
In contrast, as shown in Fig. \ref{fig:mode_open_cavity} (a) with green diamonds, in a passive open cavity the eigenfrequencies are located in the lower-half frequency plane, because of the energy leakage towards the exterior. To limit the radiation loss of the open cavity we consider that the outside medium has a permittivity of $\eps_d=3$.  
The radiation loss causes the spectrum to be shifted in the direction of the lower-half plane. The frequency shift (negative imaginary part) is approximately constant for all modes.\\
This property still holds true when the gain parameter $\eps_{xz}$ is different from zero. 
For example, for an open cavity with a real-valued $\eps_{xz}$ smaller than the gain thresholds  $\eps_{xz,\text{th}}^-<\eps_{xz}<\eps_{xz,\text{th}}^+$, all the eigenfrequencies have roughly the same imaginary part (not shown). Conversely, for $\eps_{xz}>\eps_{xz,\text{th}}^+$ or $\eps_{xz}<\eps_{xz,\text{th}}^-$ the modes split into two branches in some frequency ranges [red dots in Fig. \ref{fig:mode_open_cavity} (a)]. 
For a pure-imaginary $\eps_{xz}$ there is no threshold, and consequently the modes always split into two branches [blues crosses in Fig. \ref{fig:mode_open_cavity} (a)], similar to the closed cavity.
Interestingly, for a sufficiently large $\eps_{xz}$, some modes exhibit a negative imaginary part very close to zero, meaning that their lifetime in the cavity is very long [Fig. \ref{fig:mode_open_cavity} (b)]. 
Thus, the chiral-gain provided by the material can indeed compensate the radiation loss effects.
By finely adjusting the resonator parameters, it is possible to  achieve a state where the cavity gain compensates almost exactly the cavity leakage for the eigenmode with the largest lifetime. 
This is the scenario represented in Fig. \ref{fig:mode_open_cavity} (b), where the mode frequency is only slightly below the real-frequency axis. This can be done either by tuning the corrugation orientation (green cross) of the Kildal's mirror or the strength of the gain parameter (purple triangle). 
In this peculiar situation, that we will refer to as the “perfectly tuned” case, the field inside the cavity oscillates in time without any noticeable decay. It corresponds to a system that can store light for a long period of time, despite being open to the exterior. 
Such systems may have potential applications in optical memories \cite{alexoudi_optical_2020} or in sensing. 
By further increasing the gain coefficient $\eps_{xz}$ beyond the perfectly tuned case, the imaginary part of the eigenfrequency of at least one mode becomes positive, causing the open cavity to become an oscillator (laser). It is relevant to note that in the perfectly tuned case (lasing threshold) the field in the dielectric region corresponds to an outgoing propagating plane wave, implying that the non-Hermitian material continuously radiates energy towards the dielectric. 


\section{Lasing-Anti-Lasing Operation and the reciprocity constraint} \label{sec:CPA_laser}

In the following, we show that time-reversal invariant open-resonators relying on materials with chiral-gain can be operated in a regime where they behave as a CPA-laser \cite{longhi_mathcalpt-symmetric_2010,chong_PT-symmetry_2011,longhi_pt-symmetric_2014,wong_lasing_2016,longhi_time-reversed_2011,schackert_three-wave_2013}.
A CPA-laser simultaneously behaves as an anti-laser -- signifying that it can fully absorb an incoming signal -- and as ``quasi-laser'' signifying that it can interact resonantly with an incoming wave and store efficiently its energy with no sensible relaxation in time. Previous implementations of linear  CPA-lasers \cite{longhi_mathcalpt-symmetric_2010,chong_PT-symmetry_2011,longhi_pt-symmetric_2014,wong_lasing_2016} were based on combining different gain and lossy materials. In contrast, our system integrates both gain and loss within the same material.  We show that unlike conventional passive resonators \cite{haus_waves_1983}, the external coupling  in our resonator is not limited by reciprocity. Thereby our system can efficiently harvest a significant amount of energy, even for incoming pulses with relatively short-duration.

\subsection{Time-reversal symmetry and anti-lasing}

In recent years, there has been significant interest in  the analytical properties of the scattering matrix of time-reversal invariant systems (e.g, formed by lossless dielectrics) \cite{chong_coherent_2010,chong_PT-symmetry_2011,grigoriev_optimization_2013,baranov_coherent_2017}. In such systems, the scattering matrix is constrained by the symmetry $\db{\vec{S}}\left( \omega  \right) = \left[ \db{\vec{S}}^{ - 1}\left( \omega^\ast \right) \right]^\ast$ \cite{silveirinha_time-reversal_2019}. Thereby, the poles and zeros of $\db{\vec{S}}$ are related by complex conjugation. Note that the poles of $\db{\vec{S}}$ determine the natural frequencies of the open-system, and thereby for a passive system lie in the lower half-plane. On the other hand, the zeros of $\db{\vec{S}}$ determine the frequencies for which the system does not generate scattering. The ``zeros'' are the frequencies associated with a nontrivial null space of the scattering matrix. For a time-reversal invariant passive platform, these zeros lie in the upper-half frequency plane. We underline that the ``zeros'' do not refer to the individual elements of the scattering matrix, but rather to the zeros of the matrix determinant.  

Several works have developed interesting strategies to harvest, absorb, and store light by leveraging these analytical properties. Notably, it has been suggested that introducing a suitable amount of loss into a system may enable the displacement of a zero of $\db{\vec{S}}$ to the real-frequency axis, thereby allowing the full absorption of an incoming wave at that specific frequency with no back-scattering. This phenomenon is known as ``coherent perfect absorption'' (CPA) or ``anti-lasing'' \cite{chong_coherent_2010,wan_time-reversed_2011}.
Moreover, it has been demonstrated that the complex zeros of the scattering matrix can be utilized to store and release light in an ideally lossless cavity for an arbitrarily long time period. This mechanism, known as ``coherent virtual absorption'' \cite{baranov_coherent_2017}, relies on exciting the system with a signal that exponentially grows over time.\\
\indent Conversely, by judiciously inserting gain into an open-resonator it is possible to push one of the poles to the real-frequency axis, so that the resonator can support oscillations that do not decay in time. We refer to this mode of operation as a ``quasi-laser''.\\
\indent Furthermore, by suitably incorporating loss and gain -- such as in $\mathcal{PT}$ symmetric systems -- it is feasible to align a pole and a zero at the same real-frequency, enabling the system to function both as an anti-laser and a quasi-laser, depending on the excitation. This dual functionality defines what is known as a ``CPA-laser'' \cite{longhi_mathcalpt-symmetric_2010,chong_PT-symmetry_2011,longhi_pt-symmetric_2014,wong_lasing_2016,longhi_time-reversed_2011,schackert_three-wave_2013}. 
In the next subsections, we demonstrate that the perfectly tuned resonator introduced in Sect. \ref{sec:open_cavity} can behave as a CPA-laser.

\subsection{Analytical structure of $\db{\vec{R}}$ near a resonance}

The geometry of our cavity is the same as in Sect. \ref{sec:open_cavity}. We suppose that the permittivity tensor is real-valued to ensure that the system is time-reversal symmetric. In order to understand how the chiral-gain affects the excitation of the open cavity, next we study the analytical properties of the scattering matrix, which in our system corresponds to the reflection matrix $\db{\vec{R}}$.

In Appendix \ref{sec:derivation_Eq}, it is demonstrated that in the vicinity of a resonance ($\omega_0 = \omega'_0+i\omega''_0$) of a  open resonator with high-quality factor, the reflection matrix has the following structure:
\begin{align}
\label{Eq:Rstructure}
\db{\vec{R}} \left( \omega  \right) \approx \vec{U}^\ast \left(  \omega_0 '  \right) \cdot \begin{pmatrix}
0&\frac{A}{\left( \omega  - \omega _0 \right)}\\
\frac{1}{A^\ast}\left( \omega  - \omega _0^\ast \right)&0 \end{pmatrix} \cdot \vec{U}^\dagger\left( \omega_0 ' \right),
\end{align}
where $A$ is some constant and $\vec{U}$ is a unitary matrix. 

Consistent with the principles of time-reversal symmetry, the pole of the reflection matrix is accompanied by a complex conjugated zero, so that $|\det(\db{\vec{R}})|=1$ in the real-frequency axis, as it should be. Notably, the pole and zero pertain to \emph{distinct} elements within the middle matrix in the right-hand side of Eq. \eqref{Eq:Rstructure}. This aspect becomes particularly significant when the system gain is adjusted in such a way that $\omega_0$ approaches the real-axis. In fact, the structure of the reflection matrix preserves the influences of both the pole and zero (associated with a nontrivial nullspace), allowing them to coexist without cancellation for a perfectly tuned resonator. Note that one of the eigenvalues of the reflection matrix is controlled by the pole, whereas the second eigenvalue is controlled by the zero. 
This remarkable property is rooted on the non-Hermitian response of the material. Indeed, it is well known from the study of bound states in the continuum (BIC), that the influence of a system pole approaching the real-frequency axis through precise tuning is often completely canceled by a corresponding zero. This phenomenon renders the excitation of the bound state from an external region unfeasible, as highlighted in various studies \cite{silveirinha_trapping_2014,monticone_embedded_2014,lannebere_optical_2015,monticone_can_2019,silva_multiple_2020, prudencio_monopole_2021}. This underscores the exceptional nature of the chiral-gain response in our context, where the non-Hermitian characteristics fundamentally alter the expected analytical properties of the reflection matrix.

\begin{figure*}[!ht]
\centering
\includegraphics[width=0.9\linewidth]{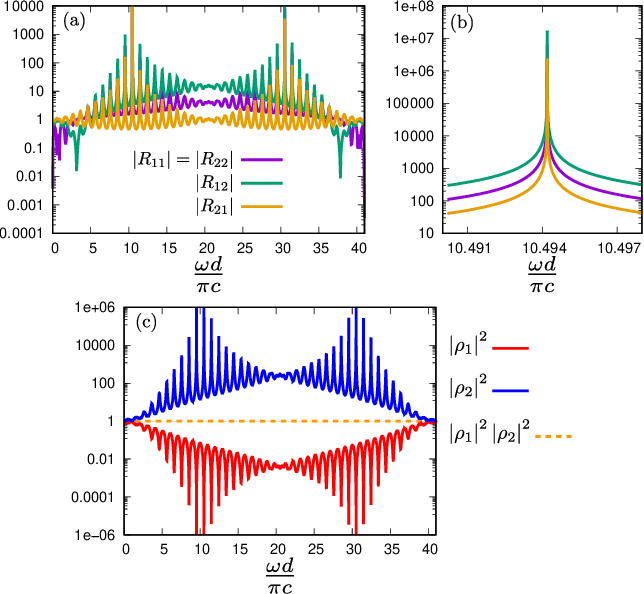}
       \caption{$\bf{(a)}$-$\bf{(b)}$ Plot of the magnitude of all entries of the reflection matrix $\db{\vec{R}}$ as a function of the normalized frequency for the perfectly tuned case of Fig. \ref{fig:mode_open_cavity} (b) (purple triangle) $\eps_{d}=3$, $\eps_{xx}=1$, $\eps_{zz}=1.1$, $\eps_{xz}=0.2001382$, $\varphi=45^\degree$. (b) Zoom in of panel (a) near the normalized frequency 10.4942. $\bf{(c)}$ Plot of the eigenvalues $|\rho_1|^2$ and $|\rho_2|^2$ of $\db{\mathcal{R}}$ as well as their product $|\rho_1|^2 |\rho_2|^2$  for the perfectly tuned cavity.
}
\label{fig:frq_excitation_open_cavity}
\end{figure*}

To illustrate the discussion, we show in Fig. \ref{fig:frq_excitation_open_cavity} the magnitude of the elements of the reflection coefficient for the perfectly tuned cavity corresponding to the purple triangle in Fig. \ref{fig:mode_open_cavity} (b). 
As seen in Fig. \ref{fig:frq_excitation_open_cavity} (a), the elements of the reflection matrix have multiple peaks whose positions match the eigenfrequencies of the open cavity. The peak associated with the almost perfectly tuned case [Fig. \ref{fig:frq_excitation_open_cavity} (b)] has a very large magnitude that can be made as large as desired by fine tuning the system parameters, in agreement with Eq. \eqref{Eq:Rstructure}. This behavior suggests that different from a perfectly tuned BIC resonator \cite{silveirinha_trapping_2014,monticone_embedded_2014,lannebere_optical_2015,monticone_can_2019}, the trapped mode of our system can be excited from the outside. This property will be analyzed in detail in the next subsection with time domain simulations. Furthermore, when the cavity is finely tuned so that $\omega_0$ approaches the real axis, the reflectivities provided by the open cavity vary as $|\rho_1|^2 \to 0$ and $|\rho_2|^2 \to \infty$, confirming that the effects of the pole and of the zero do not cancel out [see Fig. \ref{fig:frq_excitation_open_cavity} (c)].

\subsection{Time excitation of the open cavity and the reciprocity constraint} \label{sec:time_excitation}

\begin{figure*}[!ht]
\centering
\includegraphics[width=0.8\linewidth]{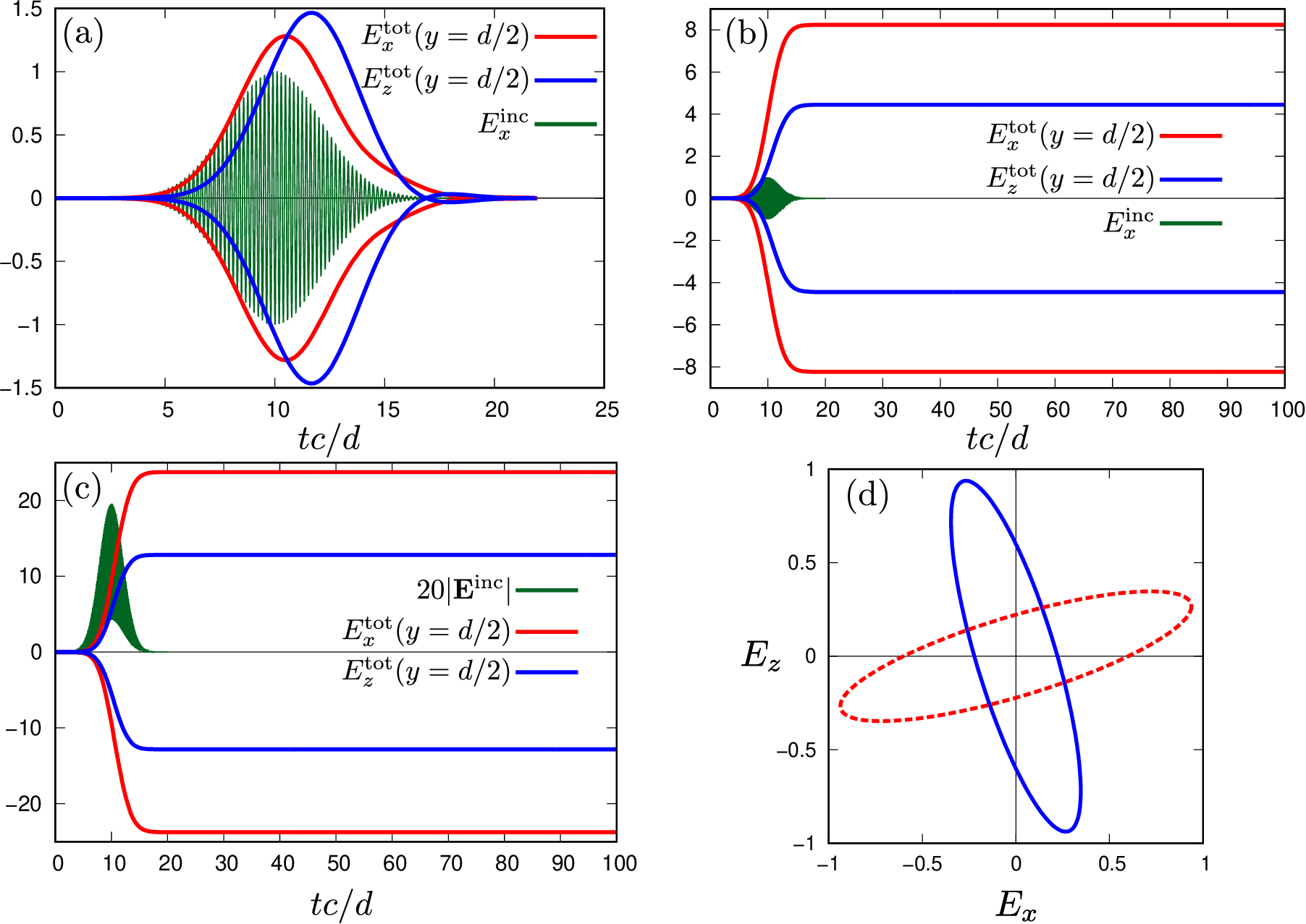}
       \caption{Time excitation of the open chiral-gain medium cavity with an incident Gaussian pulse [Eq. \eqref{E:incident_gaussian_pulse}]. $\bf{(a)}$ Time evolution of the electric field envelope in the middle of the cavity ($y=d/2$) as a function of the normalized time for $\eps_{xz}=0$, $\eps_d=3$ and with the field polarization  along $x$. $\bf{(b)}$ Similar to (a) but for the perfectly tuned case. 
       $\bf{(c)}$ Similar to (b) but for an incident field polarized with the optimal polarization [see (d)]. For clarity, the amplitude of the incident field was multiplied by a factor of 20. $\bf{(d)}$ Polarization curves for the optimal incident field (blue solid line) and for the   output (reflected) field for both (b) and (c) (dashed red line). In all the plots, the incident pulse parameters are $\sigma= 3 d/c \approx 15.74\times 2\pi/\omega_0$, $t_0=10 d/ c$ and $\omega_0\approx10.4942 \pi c /d$.}
\label{fig:time-excitation_gaussian}
\end{figure*}

Let us now analyze the excitation of a high-quality factor open cavity. Consider first a time harmonic excitation with the incident field polarized as $\vec{E}_{\text{eig},1}$. In the limit ($\omega''_0 \to 0^-$), the reflected wave is vanishingly weak because $|\rho_1| \to 0$. This field polarization is thereby completely absorbed by the resonator, corresponding to a form of coherent perfect absorption (anti-lasing). 
This effect is a consequence of the  ``zero'' of the scattering matrix ($|\rho_1| \to 0$) on the real frequency axis. It is relevant to note that under a time-reversal, the scattering problem just discussed, maps into a scenario where the resonator radiates away a plane wave with frequency $\omega'_0$ without any external excitation. This is precisely the perfectly tuned eigenstate discussed in Sect. \ref{sec:open_cavity}, corresponding to the lasing threshold.\\
Consider now, a time harmonic excitation with the incident polarization determined by $\vec{E}_{\text{eig},2}$. In this case, the response of the open-resonator diverges $|\rho_2| \to \infty$, indicating that there is no time-harmonic steady-state (quasi-laser resonance). This divergence is caused by the fact that in the limit $\omega''_0 \to 0^-$ the excitation frequency matches a pole of the scattering matrix.\\
To gain a deeper insight of the consequences of the 
time-harmonic response collapse, we conducted a numerical study of the dynamic response of the open-resonator for an  incident wave with a Gaussian profile
\begin{align} \label{E:incident_gaussian_pulse}
 \vec{E}^\text{inc}(y=0,t)=\mathrm{Re}\left\{\vec{E}_{0}^\text{inc} \e{-i\omega_0 t} \e{-\left(\frac{t-t_0}{\sigma}\right)^2} \right\}.
\end{align}
Here $\sigma$ is proportional to the full-width half-maximum (FWHM) of the Gaussian pulse, $t_0$ is
the time instant for which the incoming wave is peaked at $y=0$, 
and $\vec{E}_{0}^\text{inc}$ is a vector in the $xoz$ plane that determines the polarization of the incident field.
Using the formalism described in Appendix \ref{ApC}, we determined the time evolution of incident Gaussian pulses with
different polarizations (Fig. \ref{fig:time-excitation_gaussian}). 
To begin with, we consider a cavity with no gain ($\eps_{xz}=0$) and an incident field polarized along the $x$ direction [Fig. \ref{fig:time-excitation_gaussian} (a)]. 
We examine the behavior of the electric field envelope, calculated using Eq. \eqref{E:E_time_domain}, in the middle of the cavity. 
We see that the incident field excites a wave that is initially polarized along $x$, but that after hitting the Kildal mirror gains a component along $z$. Both components of the fields inside the cavity quickly decay after the end of the excitation, as expected from the value of the imaginary part of the eigenfrequencies ($\omega''\approx-0.664 c/d$) represented in Fig. \ref{fig:mode_open_cavity} (a).\\ 
The time evolution of the incident pulse for a perfectly tuned cavity [with the parameters corresponding to the purple triangle of Fig. \ref{fig:mode_open_cavity} (b)] is shown in Fig. \ref{fig:time-excitation_gaussian} (b). 
The duration of the incident pulse was chosen large enough (about eighty cycles of oscillation) so that only the trapped state can be excited efficiently.  Clearly, even though a zero and pole of the reflection matrix coexist at the same frequency, the incident wave is able to efficiently pump the eigenmode of the open resonator.
As before, the Kildal’s mirror is responsible for creating a field component along
the $z$-direction. Since $\eps_{xz} \neq 0$, the energy associated with the $z$-field component can be
subsequently converted back into a $x$ component. Remarkably it is observed that after the
incident pulse terminates, the field inside the cavity does not decay with time, confirming that the
trapped mode can indeed be excited by the incoming wave. The eigenstate does not
relax despite the constant field leakage towards the exterior (not shown; see also  the discussion in the end of Sect. \ref{sec:open_cavity}). In fact, the leakage through the open wall is exactly compensated by the gain material. 
Consistent with the principles of time-reversal symmetry, the polarization of the field radiated by the cavity (the eigenstate) after the excitation terminates is coincident with $\vec{E}_{\text{eig},1}^\ast$. As shown in Fig. \ref{fig:time-excitation_gaussian} (d) (dashed red line) the output polarization is elliptical and the major axis is slightly tilted with respect to the $x$ axis.\\
The magnitude of the field stored inside the resonator depends strongly on the
polarization of the incident field as seen by comparing Figs.\ref{fig:time-excitation_gaussian} (b)--(c). Indeed, our numerical study  confirms that the optimal polarization, i.e. the polarization that maximizes the energy transferred to the trapped state, is determined by $\vec{E}_{\text{eig},2}$ evaluated at the eigenstate frequency. As seen in Fig. \ref{fig:time-excitation_gaussian} (d), this polarization is elliptical, rotating clockwise, with the ellipse major axis making an angle of around 17$^\circ$ with respect to the $z$ axis. 
For this optimal case the field stored in the cavity is about 3 times larger than for a linear polarization along the $x$ axis [Fig. \ref{fig:time-excitation_gaussian} (b)], but only roughly 2\% larger than a linear polarization parallel to the major axis of the ellipse (not shown). 
Moreover, the polarization of the output field [dashed red line in Fig. \ref{fig:time-excitation_gaussian} (d)] is insensitive to the polarization of the incoming wave. This confirms that the orthogonal polarization described by $\vec{E}_{\text{eig},1}$ is fully absorbed (anti-lasing) by the cavity and hence plays no role in determining the reflected wave.\\
The magnitude of the field stored in the resonator also depends on the pulse duration. Remarkably, despite the relatively short duration of the incoming pulse in the example of Fig. \ref{fig:time-excitation_gaussian} it is feasible to inject a substantial amount of energy into the open cavity. We verified that for a passive and reciprocal dielectric resonator with the same quality factor, the field amplitude inside the cavity would be near zero (not shown), for the same excitation. This happens due to the reciprocity constraint discussed in Refs. \cite{haus_waves_1983,mann_nonreciprocal_2019} and appendix \ref{sec:constraint_input_coupling}, which imposes that the amount of energy that can be pumped into a time-reversal symmetric and passive resonator is proportional to the energy leakage towards the exterior [see Eq. \eqref{E:result_haus}]. 
Interestingly, our cavity, despite being time-reversal invariant and thus seemingly subject to the same limitations, behaves differently due to its non-passive nature. Indeed, different from standard resonators, it is demonstrated in appendix \ref{sec:constraint_input_coupling} that in our system the energy leakage towards the exterior is not controlled by the decay rate of the eigenmode, because the cavity is pumped by the material gain. Thereby, our system is not constrained by the result of \cite{haus_waves_1983}. In fact, the radiation leakage remains strong when the quality factor of the eigenstate approaches infinity due to the gain effect.

\begin{figure*}[!ht]
\centering
\includegraphics[width=0.8\linewidth]{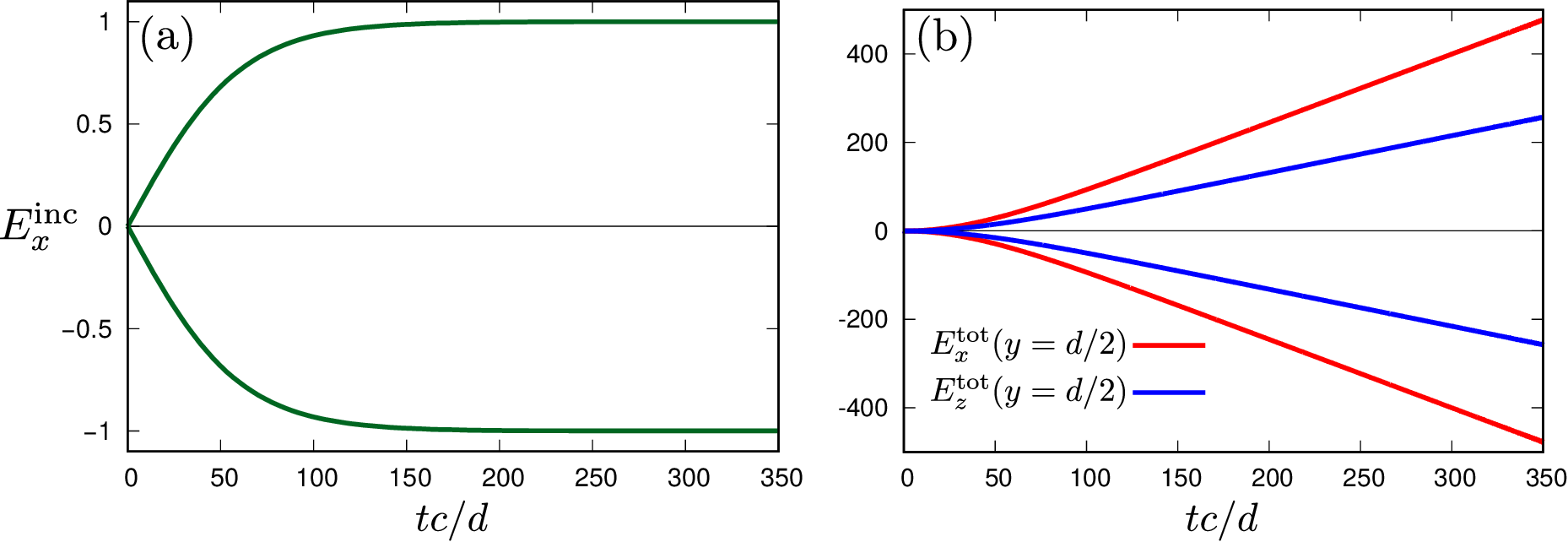}
       \caption{ Time excitation of the open-cavity with a continuous wave pulse [Eq. \eqref{E:continuous_incident_field}]  polarized along $x$. $\bf{(a)}$ Envelope of the incident field for a pulse with $\omega_0\approx10.4942 \pi c/d$, $\alpha= 60d/c$ and $\eps_d=3$. $\bf{(b)}$ Time evolution of the envelope of the electric field in the middle of the perfectly tuned cavity ($y=d / 2$) as a function of the normalized time.}
\label{fig:time-excitation_tanh}
\end{figure*}

For instance, in the perfectly tuned scenario depicted in Fig. \ref{fig:mode_open_cavity} (b), the leakage to the exterior is comparable to that in the scenario with $\eps_{xz}=0$ shown in Fig. \ref{fig:mode_open_cavity} (a), yet the quality factors in these two cases are dramatically different. This discrepancy elucidates the ability to channel a significant amount of energy into the cavity within a limited number of oscillation cycles, thereby surpassing the reciprocity constraint. \\
It is interesting to further investigate the time dynamics of the chiral-gain open-resonator when it is subjected to a very long monochromatic excitation. To this end, in the last example we suppose that the incident pulse for $t \geq 0$ is of the form
\begin{align}\label{E:continuous_incident_field}
 \vec{E}^\text{inc}(y,t)=\mathrm{Re}\left\{\vec{E}_{0}^\text{inc} \e{i\left(\frac{\omega_0}{c}\sqrt{\eps_d}y -\omega_0 t\right)} \tanh\left(\frac{1}{\alpha}\left[t-\frac{y\sqrt{\eps_d}}{c}\right]\right) \right\}
\end{align}
where $\alpha$ determines the rising time of the pulse [see Fig. \ref{fig:time-excitation_tanh} (a)]. 
In this case, one can see from Fig. \ref{fig:time-excitation_tanh} (b) that in the stationary regime the magnitude of the field stored inside the resonator grows linearly with time. This behavior is analogous to that of a lossless LC circuit pumped at the resonance, i.e., a resonator with infinite quality factor under a time-harmonic excitation.
This divergent behavior arises due to the pole of the reflection matrix in the real-frequency axis (quasi-laser), and explains the collapse of the time-harmonic regime which predicts an infinite response.
Clearly, different from conventional resonators, the amount of energy stored into the open-cavity depends solely on the duration of the excitation, not on the amplitude of the incoming wave. In practice, the maximum amount of energy that can be pumped into the cavity will be limited by nonlinear effects.
\\
In summary, we showed that a chiral-gain cavity can be operated in a peculiar regime corresponding to the threshold of lasing/anti-lasing, similar to a CPA-laser. For an excitation polarized as $\vec{E}_{\text{eig},1}$ the incoming wave is completely absorbed by the resonator (anti-lasing operation). On the other hand, for an excitation polarized as  $\vec{E}_{\text{eig},2}$ the system exhibits a strongly resonant response due to the excitation of the cavity mode (lasing threshold). The cavity is capable of efficiently collecting and storing the energy of the incoming wave within it, even for short-duration pulses, without being limited by a reciprocity constraint on the external coupling strength. 
The field trapped inside the cavity does not decay over time. The amount of energy stored in the cavity is proportional to the duration of the excitation.


\section{Conclusions}

We investigated various applications facilitated by chiral-gain media, revealing their potential in the development of electrically pumped chiral lasers, polarization-controlled non-Hermitian mirrors, and light storage. In particular, we demonstrated that the non-Hermitian properties of chiral-gain media lay the groundwork for creating CPA-lasers using a single material that integrates both gain and loss responses.

This capability sets our approach apart from traditional resonators, such as those supporting BICs, which typically suffer from inefficient excitation due to pole-zero cancellation. By contrast, the unique analytical features of our system scattering matrix allow a zero and a pole to coexist at the same frequency. Additionally, we established that our system transcends the limitations imposed by reciprocity. As a result, our CPA-laser can be efficiently excited with signals of relatively short duration and can retain the captured energy without significant decay. Thus, chiral-gain materials emerge as a promising solution for advancing nanophotonic technologies.


\section*{Acknowledgments}
This work was partially funded by the
Institution of Engineering and Technology (IET), by the Simons Foundation  award 733700
(Simons Collaboration in Mathematics and Physics, "Harnessing
Universal Symmetry Concepts for Extreme Wave Phenomena") and by
Instituto de Telecomunica\c{c}\~{o}es under Project N$^\circ$.
UID/EEA/50008/2020. S.L. acknowledges FCT and IT-Coimbra for
the research financial support with reference DL
57/2016/CP1353/CT000. D.E.F. acknowledges financial support by IT-Lisbon and
FCT under a research contract Ref. CEECINST/00058/2021/
CP2816/CT0003. T.A.M. acknowledges FCT for research financial support with reference CEECIND/04530/2017/CP1393/CT0004 (DOI 10.54499/CEECIND/04530/2017/CP1393/CT0004) under the CEEC Individual 2017, and IT-Coimbra for the contract as an assistant researcher with reference CT/N$^\circ$. 004/2019-F00069. 


\appendix

\section{Reflection matrix} \label{sec:reflection_matrix}
In this Appendix, we derive the reflection matrix $ \db{\vec{R}}$ for the configuration of Fig.\ref{fig:open_MOSFET_cavity}.\\
We start by noting that the fields at the Kildal mirror  ($y=d$) are necessarily of the form:
\begin{align}\label{E:boundary_condition_kildal_2}
\vec{f}(d)=A_E\hat{\vec{t}}_E+A_H\hat{\vec{t}}_H
\end{align}
where $A_E$ and $A_H$ are some unknown coefficients. Using the transfer matrix [Eq. \eqref{E:solutions_f}],  the fields at $y=0$ can be written as $\vec{f}(0)=\e{i d \db{\vec{M}} }\cdot \vec{f}(d)$, or equivalently:
\begin{align}
\begin{pmatrix}E_x \\E_z \\H_x \\H_z \end{pmatrix}_{y=0}=\begin{pmatrix}
 \hat{\vec{u}}_1 \cdot \e{i d \db{\vec{M}} }\cdot \hat{\vec{t}}_E  & & \hat{\vec{u}}_1 \cdot \e{i d \db{\vec{M}} }\cdot \hat{\vec{t}}_H \\ \hat{\vec{u}}_2 \cdot \e{i d \db{\vec{M}} }\cdot \hat{\vec{t}}_E  & & \hat{\vec{u}}_2 \cdot \e{i d \db{\vec{M}} }\cdot \hat{\vec{t}}_H \\ \hat{\vec{u}}_3 \cdot \e{i d \db{\vec{M}} }\cdot \hat{\vec{t}}_E  & &\hat{\vec{u}}_3 \cdot \e{i d \db{\vec{M}} }\cdot \hat{\vec{t}}_H \\\hat{\vec{u}}_4 \cdot \e{i d \db{\vec{M}} }\cdot \hat{\vec{t}}_E  & & \hat{\vec{u}}_4 \cdot \e{i d \db{\vec{M}} }\cdot \hat{\vec{t}}_H \end{pmatrix} \cdot \begin{pmatrix}
 A_E \\ A_H \end{pmatrix} \label{E:f_zero_kildal}
\end{align}
Next, we introduce an impedance matrix $\db{\vec{Z}}$ that relates the transverse components of the electric field $\vec{E}_t$ and  magnetic  field $\vec{H}_t$ at the immediate vicinity of the dielectric interface ($y=0^+$) as \cite{morgado_single_2016,silveirinha_fluctuation_2018,latioui_lateral_2019}:
\begin{align}\label{E:def_impedance_matrix}
 \db{\vec{J}}  \cdot \vec{E}_t= \db{\vec{Z}} \cdot \vec{H}_t
\end{align}
where 
\begin{align}\label{E:matrix_J}
\db{\vec{J}}=\begin{pmatrix} 0 & 1 \\-1 & 0 \end{pmatrix}
\end{align}
In order to determine the impedance matrix, we use Eq. \eqref{E:f_zero_kildal}, to obtain
\begin{align}
\db{\vec{J}} \cdot \vec{E}_t(0)&= \db{\vec{J}} \cdot \begin{pmatrix}
 \hat{\vec{u}}_1 \cdot \e{i d \db{\vec{M}} }\cdot \hat{\vec{t}}_E  & \hat{\vec{u}}_1 \cdot \e{i d \db{\vec{M}} }\cdot \hat{\vec{t}}_H \\ \hat{\vec{u}}_2 \cdot \e{i d \db{\vec{M}} }\cdot \hat{\vec{t}}_E  & \hat{\vec{u}}_2 \cdot \e{i d \db{\vec{M}} }\cdot \hat{\vec{t}}_H \end{pmatrix} \cdot \begin{pmatrix}
 A_E \\ A_H \end{pmatrix} \nonumber \\
 &= \db{\vec{J}} \cdot \begin{pmatrix}
 \hat{\vec{u}}_1 \cdot \e{i d \db{\vec{M}} }\cdot \hat{\vec{t}}_E  & \hat{\vec{u}}_1 \cdot \e{i d \db{\vec{M}} }\cdot \hat{\vec{t}}_H \\ \hat{\vec{u}}_2 \cdot \e{i d \db{\vec{M}} }\cdot \hat{\vec{t}}_E  & \hat{\vec{u}}_2 \cdot \e{i d \db{\vec{M}} }\cdot \hat{\vec{t}}_H \end{pmatrix} \cdot \begin{pmatrix}
\hat{\vec{u}}_3 \cdot \e{i d \db{\vec{M}} }\cdot \hat{\vec{t}}_E  & \hat{\vec{u}}_3 \cdot \e{i d \db{\vec{M}} }\cdot \hat{\vec{t}}_H \\\hat{\vec{u}}_4 \cdot \e{i d \db{\vec{M}} }\cdot \hat{\vec{t}}_E  & \hat{\vec{u}}_4 \cdot \e{i d \db{\vec{M}} }\cdot \hat{\vec{t}}_H \end{pmatrix}^{-1} \cdot \begin{pmatrix}H_x \\H_z \end{pmatrix}_{y=0}.  \label{E:derivation_impedance_matrix}
\end{align}
Therefore, comparing with Eq. \eqref{E:def_impedance_matrix} we see that the impedance matrix at $y=0$ is given by:
\begin{align}
\db{\vec{Z}} &=\db{\vec{J}} \cdot \begin{pmatrix}
 \hat{\vec{u}}_1 \cdot \e{i d \db{\vec{M}} }\cdot \hat{\vec{t}}_E  & \hat{\vec{u}}_1 \cdot \e{i d \db{\vec{M}} }\cdot \hat{\vec{t}}_H \\ \hat{\vec{u}}_2 \cdot \e{i d \db{\vec{M}} }\cdot \hat{\vec{t}}_E  & \hat{\vec{u}}_2 \cdot \e{i d \db{\vec{M}} }\cdot \hat{\vec{t}}_H \end{pmatrix} \cdot  \begin{pmatrix} \hat{\vec{u}}_3 \cdot \e{i d \db{\vec{M}} }\cdot \hat{\vec{t}}_E  & \hat{\vec{u}}_3 \cdot \e{i d \db{\vec{M}} }\cdot \hat{\vec{t}}_H \\\hat{\vec{u}}_4 \cdot \e{i d \db{\vec{M}} }\cdot \hat{\vec{t}}_E  & \hat{\vec{u}}_4 \cdot \e{i d \db{\vec{M}} }\cdot \hat{\vec{t}}_H \end{pmatrix}^{-1}.
\end{align}
On the other hand, for a plane wave propagating in a dielectric with permittivity $\eps_d$ the transverse electric and magnetic fields are related as:
\begin{align}
\vec{H}_t^\pm
= \pm \frac{\sqrt{\eps_d}}{\eta_0}   \db{\vec{J}}  \cdot \vec{E}_t^\pm  \label{E:H_transverse_dielectric}
\end{align}
where the $\pm$ sign determines whether the wave propagates towards the +$y$ or the $-y$ direction and $\eta_0=\sqrt{\mu_0/\eps_0}$ is the free-space impedance.\\
Hence, using Eqs. \eqref{E:E_reflected} and \eqref{E:H_transverse_dielectric} it follows that the total field at the dielectric side of the boundary ($y=0^-$) is related to the incident field as:
\begin{align}
\vec{f}(0)
=\begin{pmatrix} \left( \db{\vec{1}}_{2\times 2} + \db{\vec{R}}\right) \cdot \vec{E}_t^\text{inc} \\  \frac{\sqrt{\eps_d}}{\eta_0}   \db{\vec{J}}  \cdot \left( \db{\vec{1}}_{2\times 2} - \db{\vec{R}}\right) \cdot \vec{E}_t^\text{inc} \end{pmatrix}
\end{align}
where $\vec{1}_{2 \times 2}$ is the $2 \times 2$ identity matrix.
At the other side of the interface ($y=0^+$), the fields satisfy \eqref{E:def_impedance_matrix}. Matching the transverse components of the electric and magnetic fields at the two sides of the boundary, it follows that:
\begin{align}\label{E:eq_matching_fields}
\db{\vec{J}}  \cdot \left( \db{\vec{1}}_{2\times 2} + \db{\vec{R}}\right) = \frac{\sqrt{\eps_d}}{\eta_0}   \db{\vec{Z}}  \cdot \db{\vec{J}}  \cdot \left( \db{\vec{1}}_{2\times 2} - \db{\vec{R}}\right) 
\end{align}
%
After some manipulations, it is readily found that the reflection coefficient matrix is given by
\begin{align} \label{E:reflection_matrix}
 \db{\vec{R}} &= \left( -\frac{\sqrt{\eps_d}}{\eta_0}  \db{\vec{J}}  \cdot \db{\vec{Z}}  \cdot \db{\vec{J}} + \db{\vec{1}}_{2\times 2}\right)^{-1} \cdot \left( -\frac{\sqrt{\eps_d}}{\eta_0}  \db{\vec{J}}  \cdot \db{\vec{Z}}  \cdot \db{\vec{J}} - \db{\vec{1}}_{2\times 2} \right)
\end{align}
The derivation assumes that the inverse matrix in \eqref{E:derivation_impedance_matrix} is well defined. 
 When it is not the case, one can use the alternative formula that uses the inverse of $\db{\vec{Z}}$: 
\begin{align} \label{E:reflection_matrix_2}
 \db{\vec{R}} &= \left( \frac{\sqrt{\eps_d}}{\eta_0} \db{\vec{1}}_{2\times 2} -\db{\vec{J}}  \cdot \db{\vec{Z}}^{-1} \cdot \db{\vec{J}}    \right)^{-1} \cdot \left( \frac{\sqrt{\eps_d}}{\eta_0} \db{\vec{1}}_{2\times 2} +\db{\vec{J}}  \cdot \db{\vec{Z}}^{-1} \cdot \db{\vec{J}}    \right).
\end{align}
The matrix $\db{\vec{Z}}^{-1}$ is typically well defined at points where $\db{\vec{Z}}$ is not. 

\section{Derivation of Eq. \eqref{Eq:Rstructure}}
\label{sec:derivation_Eq}

To characterize the analytical structure of the reflection matrix, first we note that Eq. \eqref{E:reflected_eigenmode} can be rewritten as: 
\begin{align}
\db{\vec{R}}  \cdot \vec{U}\left( \omega  \right) = \vec{U}^\ast\left( \omega  \right) \cdot \begin{pmatrix}
0&a_2\left( \omega  \right)\\ a_1\left( \omega  \right)&0 \end{pmatrix}.
\end{align}
The identity holds true in the real-frequency axis.
Here, $\vec{U}$ is a matrix whose columns are the normalized eigenvectors ($\vec{E}_{\text{eig},1}$ and $\vec{E}_{\text{eig},2}$) of the reflectance matrix ($\db{\mathcal{R}}$) and ${a_j}\left( \omega  \right) = \left| {{\rho _j}} \right|{e^{i{\theta _j}}}$. Since $\vec{U}$ is a unitary matrix and varies little near the resonance, for a high-quality resonator it is possible to write:
\begin{align}
\db{\vec{R}} \left( \omega  \right) \approx \vec{U}^\ast\left(\omega_0' \right) \cdot \begin{pmatrix}
0&a_2\left( \omega  \right)\\ a_1\left( \omega  \right)&0 \end{pmatrix} \cdot \vec{U}^\dagger \left(\omega_0 ' \right). 
\end{align}
The resonance corresponds to the pole of $a_2$, which thereby may be approximated by $a_2 \approx  A/(\omega - \omega_0)$, where $A$ is some constant. 

In order that our theory is applicable away from the real-frequency axis, we enforce
that the
 reflection matrix satisfies the time-reversal symmetry condition: $\db{\vec{R}}\left( \omega  \right) = {\left[ {{{\db{\vec{R}}}^{ - 1}}\left( {{\omega ^\ast}} \right)} \right]^\ast}$. It is easy to check that this possible only if ${a_1}\left( \omega  \right) = {1 \mathord{\left/
 {\vphantom {1 {{{\left[ {{a_2}\left( {{\omega ^\ast}} \right)} \right]}^\ast}}}} \right.
 \kern-\nulldelimiterspace} {{{\left[ {{a_2}\left( {{\omega ^\ast}} \right)} \right]}^\ast}}}$. These approximations for $a_1$ and $a_2$ yield Eq. 
\eqref{Eq:Rstructure}.

\section{Time domain formalism}
\label{ApC}

Here, we describe the formalism used to characterize the time domain electric field $\vec{E}^\text{tot}(y,t)$ in the resonator problem.

We rely on Fourier theory. The Fourier transform $\psi(\vec{r},\omega)$ of a time signal $\psi(\vec{r},t)$ is defined as $ \psi(\vec{r},\omega)=\int_{-\infty}^\infty dt \psi(\vec{r},t) \e{i\omega t}$. Then, it follows that the complex amplitude of the incident field [Eq. \eqref{E:incident_gaussian_pulse}] in the frequency domain is
\begin{align}
\vec{E}^\text{inc}(y,\omega)&=\sigma \sqrt{\pi}  \vec{E}_{0}^\text{inc}\e{ik_d y}\e{ i (\omega-\omega_0)t_0}\e{ -\left(\frac{\sigma  [\omega-\omega_0]}{2}\right)^2}    
\end{align}
where $k_d=\sqrt{\eps_d}\omega/c $. Using Eqs. \eqref{E:solutions_f}, \eqref{E:E_reflected} and \eqref{E:H_transverse_dielectric} it can be shown that in the frequency domain the total field $\vec{E}^\text{tot}$ is:
\begin{align}
\vec{E}^\text{tot}(y,\omega)&=
\begin{cases}\vec{E}^\text{inc}(y,\omega) +  \e{-i k_d y }\db{\vec{R}} \cdot \vec{E}^\text{inc}(0,\omega)  & y\leq0 \\
\left(\hat{\vec{x}} \otimes \hat{\vec{u}}_1 + \hat{\vec{z}} \otimes \hat{\vec{u}}_2
 \right)\cdot\e{-iy\vec{M}}\cdot  \begin{pmatrix}\left( \vec{1}_{2 \times 2}+\db{\vec{R}}\right)\cdot \vec{E}^\text{inc}(0,\omega) \\ \frac{\sqrt{\eps_d}}{\eta_0} \vec{J} \cdot \left( \vec{1}_{2 \times 2}-\db{\vec{R}}\right)\cdot \vec{E}^\text{inc}(0,\omega)
                                  \end{pmatrix}   & y\geq0 \\
\end{cases}
\end{align}
Note that the region $y<0$ corresponds to the dielectric region and $0<y<d$ is the chiral-gain resonator.
The field in the time domain is obtained with an inverse Fourier transformation:
\begin{align}\label{E:E_time_domain}
\vec{E}^\text{tot}(y,t)&=\mathrm{Re}\left\{\frac{1}{2\pi}\int_{-\infty}^\infty d\omega   \vec{E}^\text{tot}(y,\omega) \e{-i\omega t} \right\}.
\end{align}
For systems with spectrum not confined to the lower-half frequency plane, the Fourier transform is ill-defined and one should use instead 
\begin{align}
\vec{E}^\text{tot}( y , t )=\mathrm{Re}\left\{\frac{1}{2\pi}\e{\omega'' t} \int_{-\infty}^\infty d\omega' \vec{E}^\text{tot}(y,\omega'+i\omega'') \e{-i \omega' t} \right\}.
\end{align}
where 
$\omega''$ is chosen above all the poles of $\vec{E}^\text{tot}(y,\omega'+i\omega'')$.\\


\section{Constraint on the coupling strength of time-reversal symmetric and non-conservative resonators} \label{sec:constraint_input_coupling}
We consider a single mode resonator coupled to $n$ ports. The complex amplitude $a(t)$ of the resonator's mode is governed by  \cite{haus_waves_1983,suh_temporal_2004,mann_nonreciprocal_2019}:
\begin{align}
\frac{d a}{dt}&=-i \omega_0 a + \vec{K}\cdot \vec{s}_+
\end{align}
where $\omega_0$ is the complex resonant frequency, $\vec{s}_+=\begin{pmatrix}s_{+1} & s_{+2} & \dots & s_{+n} \end{pmatrix}^T$ is the vector describing the complex amplitude of the input waves and $\vec{K}$ the vector describing the input coupling at each port.
The complex amplitude of the outgoing waves $\vec{s}_-$ is ruled by 
\begin{align}
\vec{s}_-&=\db{\vec{c}}\cdot \vec{s}_+ + \vec{d} a
\end{align}
where $\db{\vec{c}}$ is the direct scattering matrix and $\vec{d} $ the vector describing the output coupling to the ports. The different amplitudes are normalized in such a way that the stored energy is  ${\left| a \right|^2}$, and the incoming and outgoing power fluxes are controlled by ${\left| {{s_ {+i} }} \right|^2}$ and ${\left| {{s_ {-i} }} \right|^2}$, respectively. 
\\  
For a natural mode ($\vec{s}_+=0$):
\begin{subequations}\label{E:coupled_mode_theory_natural_modes}
\begin{align}
\frac{d a}{dt}&=-i \omega_0 a \label{E:eigen_evolution_mode}\\
\vec{s}_-&= \vec{d} a \label{E:eigen_outgoing_wave}
\end{align}
\end{subequations}
and it follows that the complex amplitude of a natural mode is given by $a(t)=a_0\e{-i \omega_0 t}$ with $a_0$ a complex number.
Under a time-reversal transformation $\mathcal{T}$, $a(t)$ is transformed as
\begin{align}
\mathrm{Re}\left[a(t)\right] \xrightarrow{~\mathcal{T}~}   \mathrm{Re}\left[a(-t)\right] = \mathrm{Re}\left[a_0^\ast\e{-i \omega_0^\ast t}\right],
\end{align}
meaning that under a $\mathcal{T}$ transformation $a(t) \xrightarrow{~\mathcal{T}~} a^\ast(-t)$ . If the system exhibits time-reversal symmetry, then the time-reversed natural mode field $a^\ast(-t)=a_0^\ast\e{-i \omega_0^\ast t}$ is also solution of equations \eqref{E:coupled_mode_theory_natural_modes}  with the incoming and outgoing waves time-reversed. Under a time-reversal the incoming and outgoing waves are interchanged \cite{silveirinha_time-reversal_2019,silveirinha_hidden_2019}. Thus, there is no outgoing wave and the input wave is the time-reversal of $\vec{s}_-(t)$, given by $\vec{s}_-^\ast(-t)$. It follows that: 
\begin{align}
\frac{d a^\ast(-t)}{dt}&=-i \omega_0 a^\ast(-t) + \vec{K}\cdot \vec{s}_-^\ast(-t) \\
&=-i \omega_0 a^\ast(-t) + \vec{K}\cdot \vec{d}^\ast a^\ast(-t)
\end{align}
where we used Eq. \eqref{E:eigen_outgoing_wave}. This is equivalent to 
%
\begin{align}
i \left( \omega_0 - \omega_0^\ast \right)  a^\ast &=  \vec{K}\cdot  \vec{d}^\ast a^\ast
\end{align}
So, we have proven that for systems invariant under time-reversal symmetry:
\begin{align}\label{E:K_d_time_reversal}
 -2  \omega_0'' &=  \vec{K}\cdot  \vec{d}^\ast
\end{align}
From the definition of the radiative decay rate $\omega_\text{rad}''$, vector $\bf{d}$  satisfies \cite{haus_waves_1983,suh_temporal_2004,mann_nonreciprocal_2019}:
\begin{align}\label{E:d_radiating_decay}
 -2  \omega_\text{rad}'' &=  \vec{d}\cdot  \vec{d}^\ast
\end{align}
For one-port systems, the combination of Eqs. \eqref{E:K_d_time_reversal} and \eqref{E:d_radiating_decay} implies that the magnitude of the input coupling is:
\begin{align}
  |K|  &=    \frac{-2  \omega_0''}{\sqrt{-2  \omega_\text{rad}''}} 
\end{align}
For conservative systems, $\omega_0''=\omega_\text{rad}''$, and the previous equation reduces to the well-known result of Haus \cite{haus_waves_1983},
\begin{align}\label{E:result_haus}
|K| =\sqrt{ -2  \omega_\text{rad}''},
\end{align}
which constraints the strength of the input coupling. It can be shown that the above result remains valid for arbitrary (possibly dissipative) reciprocal systems (such systems are  constrained by $\bf{K}=\bf{d}$). 
\\
In contrast, our (two-port) non-Hermitian system is only constrained by Eqs. \eqref{E:K_d_time_reversal} and \eqref{E:d_radiating_decay}. Applying the first equation to the CPA-laser of section \ref{sec:CPA_laser} ($\omega_0''\to 0$), it follows that:
\begin{align}
  \vec{K}\cdot  \vec{d}^\ast  &=  0 \\
   -2  \omega_\text{rad}'' &=  \vec{d}\cdot  \vec{d}^\ast
\end{align}
Since the radiative decay rate does not vanish ($\vec{d}\neq 0$), it follows that $\vec{K}$ is  perpendicular to $\vec{d}$. For the problem of section \ref{sec:time_excitation}, $\vec{d} \sim \vec{E}_{\text{eig},1}^\ast$ , which corresponds to the field radiated by the natural mode of the cavity (time-reversal of the mode that is fully absorbed). This means that $\vec{K} \sim \vec{E}_{\text{eig},2}^\ast$, as expected, so that the optimal incident wave to excite the cavity is $\vec{s}_+ \sim \vec{E}_{\text{eig},2}$. The amplitude of $\vec{K}$ is not directly constrained by time-reversal symmetry.

\section*{References}
\bibliographystyle{apsrev4-1}
\bibliography{Biblio_CLEAN}

\begin{thebibliography}{49}%
\makeatletter
\providecommand \@ifxundefined [1]{%
 \@ifx{#1\undefined}
}%
\providecommand \@ifnum [1]{%
 \ifnum #1\expandafter \@firstoftwo
 \else \expandafter \@secondoftwo
 \fi
}%
\providecommand \@ifx [1]{%
 \ifx #1\expandafter \@firstoftwo
 \else \expandafter \@secondoftwo
 \fi
}%
\providecommand \natexlab [1]{#1}%
\providecommand \enquote  [1]{``#1''}%
\providecommand \bibnamefont  [1]{#1}%
\providecommand \bibfnamefont [1]{#1}%
\providecommand \citenamefont [1]{#1}%
\providecommand \href@noop [0]{\@secondoftwo}%
\providecommand \href [0]{\begingroup \@sanitize@url \@href}%
\providecommand \@href[1]{\@@startlink{#1}\@@href}%
\providecommand \@@href[1]{\endgroup#1\@@endlink}%
\providecommand \@sanitize@url [0]{\catcode `\\12\catcode `\$12\catcode
  `\&12\catcode `\#12\catcode `\^12\catcode `\_12\catcode `\%12\relax}%
\providecommand \@@startlink[1]{}%
\providecommand \@@endlink[0]{}%
\providecommand \url  [0]{\begingroup\@sanitize@url \@url }%
\providecommand \@url [1]{\endgroup\@href {#1}{\urlprefix }}%
\providecommand \urlprefix  [0]{URL }%
\providecommand \Eprint [0]{\href }%
\providecommand \doibase [0]{http://dx.doi.org/}%
\providecommand \selectlanguage [0]{\@gobble}%
\providecommand \bibinfo  [0]{\@secondoftwo}%
\providecommand \bibfield  [0]{\@secondoftwo}%
\providecommand \translation [1]{[#1]}%
\providecommand \BibitemOpen [0]{}%
\providecommand \bibitemStop [0]{}%
\providecommand \bibitemNoStop [0]{.\EOS\space}%
\providecommand \EOS [0]{\spacefactor3000\relax}%
\providecommand \BibitemShut  [1]{\csname bibitem#1\endcsname}%
\let\auto@bib@innerbib\@empty
\bibitem [{\citenamefont {Chrostowski}\ and\ \citenamefont
  {Hochberg}(2015)}]{chrostowski_silicon_2015}%
  \BibitemOpen
  \bibfield  {author} {\bibinfo {author} {\bibfnamefont {L.}~\bibnamefont
  {Chrostowski}}\ and\ \bibinfo {author} {\bibfnamefont {M.}~\bibnamefont
  {Hochberg}},\ }\href@noop {} {\emph {\bibinfo {title} {Silicon {Photonics}
  {Design}}}}\ (\bibinfo  {publisher} {Cambridge University Press},\ \bibinfo
  {year} {2015})\BibitemShut {NoStop}%
\bibitem [{\citenamefont {Komljenovic}\ \emph {et~al.}(2016)\citenamefont
  {Komljenovic}, \citenamefont {Davenport}, \citenamefont {Hulme},
  \citenamefont {Liu}, \citenamefont {Santis}, \citenamefont {Spott},
  \citenamefont {Srinivasan}, \citenamefont {Stanton}, \citenamefont {Zhang},\
  and\ \citenamefont {Bowers}}]{komljenovic_heterogeneous_2016}%
  \BibitemOpen
  \bibfield  {author} {\bibinfo {author} {\bibfnamefont {T.}~\bibnamefont
  {Komljenovic}}, \bibinfo {author} {\bibfnamefont {M.}~\bibnamefont
  {Davenport}}, \bibinfo {author} {\bibfnamefont {J.}~\bibnamefont {Hulme}},
  \bibinfo {author} {\bibfnamefont {A.~Y.}\ \bibnamefont {Liu}}, \bibinfo
  {author} {\bibfnamefont {C.~T.}\ \bibnamefont {Santis}}, \bibinfo {author}
  {\bibfnamefont {A.}~\bibnamefont {Spott}}, \bibinfo {author} {\bibfnamefont
  {S.}~\bibnamefont {Srinivasan}}, \bibinfo {author} {\bibfnamefont {E.~J.}\
  \bibnamefont {Stanton}}, \bibinfo {author} {\bibfnamefont {C.}~\bibnamefont
  {Zhang}}, \ and\ \bibinfo {author} {\bibfnamefont {J.~E.}\ \bibnamefont
  {Bowers}},\ }\href@noop {} {\bibfield  {journal} {\bibinfo  {journal} {J.
  Lightwave Technol., JLT}\ }\textbf {\bibinfo {volume} {34}},\ \bibinfo
  {pages} {20} (\bibinfo {year} {2016})}\BibitemShut {NoStop}%
\bibitem [{\citenamefont {Karabchevsky}\ \emph {et~al.}(2020)\citenamefont
  {Karabchevsky}, \citenamefont {Katiyi}, \citenamefont {Ang},\ and\
  \citenamefont {Hazan}}]{karabchevsky_chip_2020}%
  \BibitemOpen
  \bibfield  {author} {\bibinfo {author} {\bibfnamefont {A.}~\bibnamefont
  {Karabchevsky}}, \bibinfo {author} {\bibfnamefont {A.}~\bibnamefont
  {Katiyi}}, \bibinfo {author} {\bibfnamefont {A.~S.}\ \bibnamefont {Ang}}, \
  and\ \bibinfo {author} {\bibfnamefont {A.}~\bibnamefont {Hazan}},\ }\href
  {\doibase 10.1515/nanoph-2020-0204} {\bibfield  {journal} {\bibinfo
  {journal} {Nanophotonics}\ }\textbf {\bibinfo {volume} {9}},\ \bibinfo
  {pages} {3733} (\bibinfo {year} {2020})}\BibitemShut {NoStop}%
\bibitem [{\citenamefont {Zhou}\ \emph {et~al.}(2015)\citenamefont {Zhou},
  \citenamefont {Yin},\ and\ \citenamefont {Michel}}]{zhou_chip_2015}%
  \BibitemOpen
  \bibfield  {author} {\bibinfo {author} {\bibfnamefont {Z.}~\bibnamefont
  {Zhou}}, \bibinfo {author} {\bibfnamefont {B.}~\bibnamefont {Yin}}, \ and\
  \bibinfo {author} {\bibfnamefont {J.}~\bibnamefont {Michel}},\ }\href
  {\doibase 10.1038/lsa.2015.131} {\bibfield  {journal} {\bibinfo  {journal}
  {Light Sci Appl}\ }\textbf {\bibinfo {volume} {4}},\ \bibinfo {pages} {e358}
  (\bibinfo {year} {2015})}\BibitemShut {NoStop}%
\bibitem [{\citenamefont {Wang}\ \emph {et~al.}(2017)\citenamefont {Wang},
  \citenamefont {Abbasi}, \citenamefont {Dave}, \citenamefont {De~Groote},
  \citenamefont {Kumari}, \citenamefont {Kunert}, \citenamefont {Merckling},
  \citenamefont {Pantouvaki}, \citenamefont {Shi}, \citenamefont {Tian},
  \citenamefont {Van~Gasse}, \citenamefont {Verbist}, \citenamefont {Wang},
  \citenamefont {Xie}, \citenamefont {Zhang}, \citenamefont {Zhu},
  \citenamefont {Bauwelinck}, \citenamefont {Yin}, \citenamefont {Hens},
  \citenamefont {Van~Campenhout}, \citenamefont {Kuyken}, \citenamefont
  {Baets}, \citenamefont {Morthier}, \citenamefont {Van~Thourhout},\ and\
  \citenamefont {Roelkens}}]{wang_novel_2017}%
  \BibitemOpen
  \bibfield  {author} {\bibinfo {author} {\bibfnamefont {Z.}~\bibnamefont
  {Wang}}, \bibinfo {author} {\bibfnamefont {A.}~\bibnamefont {Abbasi}},
  \bibinfo {author} {\bibfnamefont {U.}~\bibnamefont {Dave}}, \bibinfo {author}
  {\bibfnamefont {A.}~\bibnamefont {De~Groote}}, \bibinfo {author}
  {\bibfnamefont {S.}~\bibnamefont {Kumari}}, \bibinfo {author} {\bibfnamefont
  {B.}~\bibnamefont {Kunert}}, \bibinfo {author} {\bibfnamefont
  {C.}~\bibnamefont {Merckling}}, \bibinfo {author} {\bibfnamefont
  {M.}~\bibnamefont {Pantouvaki}}, \bibinfo {author} {\bibfnamefont
  {Y.}~\bibnamefont {Shi}}, \bibinfo {author} {\bibfnamefont {B.}~\bibnamefont
  {Tian}}, \bibinfo {author} {\bibfnamefont {K.}~\bibnamefont {Van~Gasse}},
  \bibinfo {author} {\bibfnamefont {J.}~\bibnamefont {Verbist}}, \bibinfo
  {author} {\bibfnamefont {R.}~\bibnamefont {Wang}}, \bibinfo {author}
  {\bibfnamefont {W.}~\bibnamefont {Xie}}, \bibinfo {author} {\bibfnamefont
  {J.}~\bibnamefont {Zhang}}, \bibinfo {author} {\bibfnamefont
  {Y.}~\bibnamefont {Zhu}}, \bibinfo {author} {\bibfnamefont {J.}~\bibnamefont
  {Bauwelinck}}, \bibinfo {author} {\bibfnamefont {X.}~\bibnamefont {Yin}},
  \bibinfo {author} {\bibfnamefont {Z.}~\bibnamefont {Hens}}, \bibinfo {author}
  {\bibfnamefont {J.}~\bibnamefont {Van~Campenhout}}, \bibinfo {author}
  {\bibfnamefont {B.}~\bibnamefont {Kuyken}}, \bibinfo {author} {\bibfnamefont
  {R.}~\bibnamefont {Baets}}, \bibinfo {author} {\bibfnamefont
  {G.}~\bibnamefont {Morthier}}, \bibinfo {author} {\bibfnamefont
  {D.}~\bibnamefont {Van~Thourhout}}, \ and\ \bibinfo {author} {\bibfnamefont
  {G.}~\bibnamefont {Roelkens}},\ }\href {\doibase 10.1002/lpor.201700063}
  {\bibfield  {journal} {\bibinfo  {journal} {Laser \& Photonics Reviews}\
  }\textbf {\bibinfo {volume} {11}},\ \bibinfo {pages} {1700063} (\bibinfo
  {year} {2017})}\BibitemShut {NoStop}%
\bibitem [{\citenamefont {Zhou}\ \emph {et~al.}(2023)\citenamefont {Zhou},
  \citenamefont {Ou}, \citenamefont {Fang}, \citenamefont {Alkhazraji},
  \citenamefont {Xu}, \citenamefont {Wan},\ and\ \citenamefont
  {Bowers}}]{zhou_prospects_2023}%
  \BibitemOpen
  \bibfield  {author} {\bibinfo {author} {\bibfnamefont {Z.}~\bibnamefont
  {Zhou}}, \bibinfo {author} {\bibfnamefont {X.}~\bibnamefont {Ou}}, \bibinfo
  {author} {\bibfnamefont {Y.}~\bibnamefont {Fang}}, \bibinfo {author}
  {\bibfnamefont {E.}~\bibnamefont {Alkhazraji}}, \bibinfo {author}
  {\bibfnamefont {R.}~\bibnamefont {Xu}}, \bibinfo {author} {\bibfnamefont
  {Y.}~\bibnamefont {Wan}}, \ and\ \bibinfo {author} {\bibfnamefont {J.~E.}\
  \bibnamefont {Bowers}},\ }\href {\doibase 10.1186/s43593-022-00027-x}
  {\bibfield  {journal} {\bibinfo  {journal} {eLight}\ }\textbf {\bibinfo
  {volume} {3}},\ \bibinfo {pages} {1} (\bibinfo {year} {2023})}\BibitemShut
  {NoStop}%
\bibitem [{\citenamefont {Azzam}\ \emph {et~al.}(2020)\citenamefont {Azzam},
  \citenamefont {Kildishev}, \citenamefont {Ma}, \citenamefont {Ning},
  \citenamefont {Oulton}, \citenamefont {Shalaev}, \citenamefont {Stockman},
  \citenamefont {Xu},\ and\ \citenamefont {Zhang}}]{azzam_ten_2020}%
  \BibitemOpen
  \bibfield  {author} {\bibinfo {author} {\bibfnamefont {S.~I.}\ \bibnamefont
  {Azzam}}, \bibinfo {author} {\bibfnamefont {A.~V.}\ \bibnamefont
  {Kildishev}}, \bibinfo {author} {\bibfnamefont {R.-M.}\ \bibnamefont {Ma}},
  \bibinfo {author} {\bibfnamefont {C.-Z.}\ \bibnamefont {Ning}}, \bibinfo
  {author} {\bibfnamefont {R.}~\bibnamefont {Oulton}}, \bibinfo {author}
  {\bibfnamefont {V.~M.}\ \bibnamefont {Shalaev}}, \bibinfo {author}
  {\bibfnamefont {M.~I.}\ \bibnamefont {Stockman}}, \bibinfo {author}
  {\bibfnamefont {J.-L.}\ \bibnamefont {Xu}}, \ and\ \bibinfo {author}
  {\bibfnamefont {X.}~\bibnamefont {Zhang}},\ }\href {\doibase
  10.1038/s41377-020-0319-7} {\bibfield  {journal} {\bibinfo  {journal} {Light
  Sci Appl}\ }\textbf {\bibinfo {volume} {9}},\ \bibinfo {pages} {90} (\bibinfo
  {year} {2020})}\BibitemShut {NoStop}%
\bibitem [{\citenamefont {Levy}(2002)}]{levy_chip_2002}%
  \BibitemOpen
  \bibfield  {author} {\bibinfo {author} {\bibfnamefont {M.}~\bibnamefont
  {Levy}},\ }\href {\doibase 10.1109/JSTQE.2002.806691} {\bibfield  {journal}
  {\bibinfo  {journal} {IEEE Journal of Selected Topics in Quantum
  Electronics}\ }\textbf {\bibinfo {volume} {8}},\ \bibinfo {pages} {1300}
  (\bibinfo {year} {2002})}\BibitemShut {NoStop}%
\bibitem [{\citenamefont {Caloz}\ \emph {et~al.}(2018)\citenamefont {Caloz},
  \citenamefont {Al{\`u}}, \citenamefont {Tretyakov}, \citenamefont {Sounas},
  \citenamefont {Achouri},\ and\ \citenamefont
  {Deck-L{\'e}ger}}]{caloz_electromagnetic_2018}%
  \BibitemOpen
  \bibfield  {author} {\bibinfo {author} {\bibfnamefont {C.}~\bibnamefont
  {Caloz}}, \bibinfo {author} {\bibfnamefont {A.}~\bibnamefont {Al{\`u}}},
  \bibinfo {author} {\bibfnamefont {S.}~\bibnamefont {Tretyakov}}, \bibinfo
  {author} {\bibfnamefont {D.}~\bibnamefont {Sounas}}, \bibinfo {author}
  {\bibfnamefont {K.}~\bibnamefont {Achouri}}, \ and\ \bibinfo {author}
  {\bibfnamefont {Z.-L.}\ \bibnamefont {Deck-L{\'e}ger}},\ }\href {\doibase
  10.1103/PhysRevApplied.10.047001} {\bibfield  {journal} {\bibinfo  {journal}
  {Phys. Rev. Applied}\ }\textbf {\bibinfo {volume} {10}},\ \bibinfo {pages}
  {047001} (\bibinfo {year} {2018})}\BibitemShut {NoStop}%
\bibitem [{\citenamefont {Asadchy}\ \emph {et~al.}(2020)\citenamefont
  {Asadchy}, \citenamefont {Mirmoosa}, \citenamefont {D{\'i}az-Rubio},
  \citenamefont {Fan},\ and\ \citenamefont
  {Tretyakov}}]{asadchy_tutorial_2020}%
  \BibitemOpen
  \bibfield  {author} {\bibinfo {author} {\bibfnamefont {V.~S.}\ \bibnamefont
  {Asadchy}}, \bibinfo {author} {\bibfnamefont {M.~S.}\ \bibnamefont
  {Mirmoosa}}, \bibinfo {author} {\bibfnamefont {A.}~\bibnamefont
  {D{\'i}az-Rubio}}, \bibinfo {author} {\bibfnamefont {S.}~\bibnamefont {Fan}},
  \ and\ \bibinfo {author} {\bibfnamefont {S.~A.}\ \bibnamefont {Tretyakov}},\
  }\href {\doibase 10.1109/JPROC.2020.3012381} {\bibfield  {journal} {\bibinfo
  {journal} {Proceedings of the IEEE}\ }\textbf {\bibinfo {volume} {108}},\
  \bibinfo {pages} {1684} (\bibinfo {year} {2020})}\BibitemShut {NoStop}%
\bibitem [{\citenamefont {Lanneb{\`e}re}\ \emph {et~al.}(2022)\citenamefont
  {Lanneb{\`e}re}, \citenamefont {Fernandes}, \citenamefont {Morgado},\ and\
  \citenamefont {Silveirinha}}]{lannebere_nonreciprocal_2022}%
  \BibitemOpen
  \bibfield  {author} {\bibinfo {author} {\bibfnamefont {S.}~\bibnamefont
  {Lanneb{\`e}re}}, \bibinfo {author} {\bibfnamefont {D.~E.}\ \bibnamefont
  {Fernandes}}, \bibinfo {author} {\bibfnamefont {T.~A.}\ \bibnamefont
  {Morgado}}, \ and\ \bibinfo {author} {\bibfnamefont {M.~G.}\ \bibnamefont
  {Silveirinha}},\ }\href {\doibase 10.1103/PhysRevLett.128.013902} {\bibfield
  {journal} {\bibinfo  {journal} {Phys. Rev. Lett.}\ }\textbf {\bibinfo
  {volume} {128}},\ \bibinfo {pages} {013902} (\bibinfo {year}
  {2022})}\BibitemShut {NoStop}%
\bibitem [{\citenamefont {Fernandes}\ \emph {et~al.}(2024)\citenamefont
  {Fernandes}, \citenamefont {Lannebère}, \citenamefont {Morgado},\ and\
  \citenamefont {Silveirinha1}}]{fernandes_exceptional_2024}%
  \BibitemOpen
  \bibfield  {author} {\bibinfo {author} {\bibfnamefont {D.~E.}\ \bibnamefont
  {Fernandes}}, \bibinfo {author} {\bibfnamefont {S.}~\bibnamefont
  {Lannebère}}, \bibinfo {author} {\bibfnamefont {T.~A.}\ \bibnamefont
  {Morgado}}, \ and\ \bibinfo {author} {\bibfnamefont {M.~G.}\ \bibnamefont
  {Silveirinha1}},\ }\href@noop {} {\enquote {\bibinfo {title} {Exceptional
  points in transistor-metamaterial inspired transmission lines},}\ } (\bibinfo
  {year} {2024}),\ \Eprint {http://arxiv.org/abs/2402.03320} {arXiv:2402.03320}
  \BibitemShut {NoStop}%
\bibitem [{\citenamefont {Rappoport}\ \emph {et~al.}(2023)\citenamefont
  {Rappoport}, \citenamefont {Morgado}, \citenamefont {Lannebère},\ and\
  \citenamefont {Silveirinha}}]{rappoport_engineering_2023}%
  \BibitemOpen
  \bibfield  {author} {\bibinfo {author} {\bibfnamefont {T.~G.}\ \bibnamefont
  {Rappoport}}, \bibinfo {author} {\bibfnamefont {T.~A.}\ \bibnamefont
  {Morgado}}, \bibinfo {author} {\bibfnamefont {S.}~\bibnamefont {Lannebère}},
  \ and\ \bibinfo {author} {\bibfnamefont {M.~G.}\ \bibnamefont
  {Silveirinha}},\ }\href {\doibase 10.1103/PhysRevLett.130.076901} {\bibfield
  {journal} {\bibinfo  {journal} {Phys. Rev. Lett.}\ }\textbf {\bibinfo
  {volume} {130}},\ \bibinfo {pages} {076901} (\bibinfo {year}
  {2023})}\BibitemShut {NoStop}%
\bibitem [{\citenamefont {Morgado}\ \emph {et~al.}(2024)\citenamefont
  {Morgado}, \citenamefont {Rappoport}, \citenamefont {Tsirkin}, \citenamefont
  {Lannebère}, \citenamefont {Souza},\ and\ \citenamefont
  {Silveirinha}}]{morgado_nonhermitian_2024}%
  \BibitemOpen
  \bibfield  {author} {\bibinfo {author} {\bibfnamefont {T.~A.}\ \bibnamefont
  {Morgado}}, \bibinfo {author} {\bibfnamefont {T.~G.}\ \bibnamefont
  {Rappoport}}, \bibinfo {author} {\bibfnamefont {S.~S.}\ \bibnamefont
  {Tsirkin}}, \bibinfo {author} {\bibfnamefont {S.}~\bibnamefont {Lannebère}},
  \bibinfo {author} {\bibfnamefont {I.}~\bibnamefont {Souza}}, \ and\ \bibinfo
  {author} {\bibfnamefont {M.~G.}\ \bibnamefont {Silveirinha}},\ }\href
  {\doibase 10.48550/arXiv.2401.13764} {\enquote {\bibinfo {title}
  {Non-{Hermitian} {Linear} {Electrooptic} {Effect} in {3D} materials},}\ }
  (\bibinfo {year} {2024}),\ \Eprint {http://arxiv.org/abs/2401.13764}
  {arXiv:2401.13764} \BibitemShut {NoStop}%
\bibitem [{\citenamefont {Hakimi}\ \emph {et~al.}(2023)\citenamefont {Hakimi},
  \citenamefont {Rouhi}, \citenamefont {Rappoport}, \citenamefont
  {Silveirinha},\ and\ \citenamefont {Capolino}}]{hakimi_chiral_2023}%
  \BibitemOpen
  \bibfield  {author} {\bibinfo {author} {\bibfnamefont {A.}~\bibnamefont
  {Hakimi}}, \bibinfo {author} {\bibfnamefont {K.}~\bibnamefont {Rouhi}},
  \bibinfo {author} {\bibfnamefont {T.~G.}\ \bibnamefont {Rappoport}}, \bibinfo
  {author} {\bibfnamefont {M.~G.}\ \bibnamefont {Silveirinha}}, \ and\ \bibinfo
  {author} {\bibfnamefont {F.}~\bibnamefont {Capolino}},\ }\href {\doibase
  10.48550/arXiv.2312.15142} {\enquote {\bibinfo {title} {Chiral terahertz
  lasing with {Berry} curvature dipoles},}\ } (\bibinfo {year}
  {2023})\BibitemShut {NoStop}%
\bibitem [{\citenamefont {Altman}\ and\ \citenamefont
  {Suchy}(2011)}]{altman_reciprocity_2011}%
  \BibitemOpen
  \bibfield  {author} {\bibinfo {author} {\bibfnamefont {C.}~\bibnamefont
  {Altman}}\ and\ \bibinfo {author} {\bibfnamefont {K.}~\bibnamefont {Suchy}},\
  }\href@noop {} {\emph {\bibinfo {title} {Reciprocity, {Spatial} {Mapping} and
  {Time} {Reversal} in {Electromagnetics}}}},\ \bibinfo {edition} {2nd}\ ed.\
  (\bibinfo  {publisher} {Springer},\ \bibinfo {year} {2011})\BibitemShut
  {NoStop}%
\bibitem [{\citenamefont {Chong}\ \emph {et~al.}(2010)\citenamefont {Chong},
  \citenamefont {Ge}, \citenamefont {Cao},\ and\ \citenamefont
  {Stone}}]{chong_coherent_2010}%
  \BibitemOpen
  \bibfield  {author} {\bibinfo {author} {\bibfnamefont {Y.~D.}\ \bibnamefont
  {Chong}}, \bibinfo {author} {\bibfnamefont {L.}~\bibnamefont {Ge}}, \bibinfo
  {author} {\bibfnamefont {H.}~\bibnamefont {Cao}}, \ and\ \bibinfo {author}
  {\bibfnamefont {A.~D.}\ \bibnamefont {Stone}},\ }\href {\doibase
  10.1103/PhysRevLett.105.053901} {\bibfield  {journal} {\bibinfo  {journal}
  {Phys. Rev. Lett.}\ }\textbf {\bibinfo {volume} {105}},\ \bibinfo {pages}
  {053901} (\bibinfo {year} {2010})}\BibitemShut {NoStop}%
\bibitem [{\citenamefont {Wan}\ \emph {et~al.}(2011)\citenamefont {Wan},
  \citenamefont {Chong}, \citenamefont {Ge}, \citenamefont {Noh}, \citenamefont
  {Stone},\ and\ \citenamefont {Cao}}]{wan_time-reversed_2011}%
  \BibitemOpen
  \bibfield  {author} {\bibinfo {author} {\bibfnamefont {W.}~\bibnamefont
  {Wan}}, \bibinfo {author} {\bibfnamefont {Y.}~\bibnamefont {Chong}}, \bibinfo
  {author} {\bibfnamefont {L.}~\bibnamefont {Ge}}, \bibinfo {author}
  {\bibfnamefont {H.}~\bibnamefont {Noh}}, \bibinfo {author} {\bibfnamefont
  {A.~D.}\ \bibnamefont {Stone}}, \ and\ \bibinfo {author} {\bibfnamefont
  {H.}~\bibnamefont {Cao}},\ }\href {\doibase 10.1126/science.1200735}
  {\bibfield  {journal} {\bibinfo  {journal} {Science}\ }\textbf {\bibinfo
  {volume} {331}},\ \bibinfo {pages} {889} (\bibinfo {year}
  {2011})}\BibitemShut {NoStop}%
\bibitem [{\citenamefont {Longhi}(2010)}]{longhi_mathcalpt-symmetric_2010}%
  \BibitemOpen
  \bibfield  {author} {\bibinfo {author} {\bibfnamefont {S.}~\bibnamefont
  {Longhi}},\ }\href {\doibase 10.1103/PhysRevA.82.031801} {\bibfield
  {journal} {\bibinfo  {journal} {Phys. Rev. A}\ }\textbf {\bibinfo {volume}
  {82}},\ \bibinfo {pages} {031801} (\bibinfo {year} {2010})}\BibitemShut
  {NoStop}%
\bibitem [{\citenamefont {Chong}\ \emph {et~al.}(2011)\citenamefont {Chong},
  \citenamefont {Ge},\ and\ \citenamefont {Stone}}]{chong_PT-symmetry_2011}%
  \BibitemOpen
  \bibfield  {author} {\bibinfo {author} {\bibfnamefont {Y.~D.}\ \bibnamefont
  {Chong}}, \bibinfo {author} {\bibfnamefont {L.}~\bibnamefont {Ge}}, \ and\
  \bibinfo {author} {\bibfnamefont {A.~D.}\ \bibnamefont {Stone}},\ }\href
  {\doibase 10.1103/PhysRevLett.106.093902} {\bibfield  {journal} {\bibinfo
  {journal} {Phys. Rev. Lett.}\ }\textbf {\bibinfo {volume} {106}},\ \bibinfo
  {pages} {093902} (\bibinfo {year} {2011})}\BibitemShut {NoStop}%
\bibitem [{\citenamefont {Longhi}\ and\ \citenamefont
  {Feng}(2014)}]{longhi_pt-symmetric_2014}%
  \BibitemOpen
  \bibfield  {author} {\bibinfo {author} {\bibfnamefont {S.}~\bibnamefont
  {Longhi}}\ and\ \bibinfo {author} {\bibfnamefont {L.}~\bibnamefont {Feng}},\
  }\href {\doibase 10.1364/OL.39.005026} {\bibfield  {journal} {\bibinfo
  {journal} {Opt. Lett., OL}\ }\textbf {\bibinfo {volume} {39}},\ \bibinfo
  {pages} {5026} (\bibinfo {year} {2014})}\BibitemShut {NoStop}%
\bibitem [{\citenamefont {Wong}\ \emph {et~al.}(2016)\citenamefont {Wong},
  \citenamefont {Xu}, \citenamefont {Kim}, \citenamefont {O'Brien},
  \citenamefont {Wang}, \citenamefont {Feng},\ and\ \citenamefont
  {Zhang}}]{wong_lasing_2016}%
  \BibitemOpen
  \bibfield  {author} {\bibinfo {author} {\bibfnamefont {Z.~J.}\ \bibnamefont
  {Wong}}, \bibinfo {author} {\bibfnamefont {Y.-L.}\ \bibnamefont {Xu}},
  \bibinfo {author} {\bibfnamefont {J.}~\bibnamefont {Kim}}, \bibinfo {author}
  {\bibfnamefont {K.}~\bibnamefont {O'Brien}}, \bibinfo {author} {\bibfnamefont
  {Y.}~\bibnamefont {Wang}}, \bibinfo {author} {\bibfnamefont {L.}~\bibnamefont
  {Feng}}, \ and\ \bibinfo {author} {\bibfnamefont {X.}~\bibnamefont {Zhang}},\
  }\href {\doibase 10.1038/nphoton.2016.216} {\bibfield  {journal} {\bibinfo
  {journal} {Nature Photon}\ }\textbf {\bibinfo {volume} {10}},\ \bibinfo
  {pages} {796} (\bibinfo {year} {2016})}\BibitemShut {NoStop}%
\bibitem [{\citenamefont {Longhi}(2011)}]{longhi_time-reversed_2011}%
  \BibitemOpen
  \bibfield  {author} {\bibinfo {author} {\bibfnamefont {S.}~\bibnamefont
  {Longhi}},\ }\href {\doibase 10.1103/PhysRevLett.107.033901} {\bibfield
  {journal} {\bibinfo  {journal} {Phys. Rev. Lett.}\ }\textbf {\bibinfo
  {volume} {107}},\ \bibinfo {pages} {033901} (\bibinfo {year}
  {2011})}\BibitemShut {NoStop}%
\bibitem [{\citenamefont {Schackert}\ \emph {et~al.}(2013)\citenamefont
  {Schackert}, \citenamefont {Roy}, \citenamefont {Hatridge}, \citenamefont
  {Devoret},\ and\ \citenamefont {Stone}}]{schackert_three-wave_2013}%
  \BibitemOpen
  \bibfield  {author} {\bibinfo {author} {\bibfnamefont {F.}~\bibnamefont
  {Schackert}}, \bibinfo {author} {\bibfnamefont {A.}~\bibnamefont {Roy}},
  \bibinfo {author} {\bibfnamefont {M.}~\bibnamefont {Hatridge}}, \bibinfo
  {author} {\bibfnamefont {M.~H.}\ \bibnamefont {Devoret}}, \ and\ \bibinfo
  {author} {\bibfnamefont {A.~D.}\ \bibnamefont {Stone}},\ }\href {\doibase
  10.1103/PhysRevLett.111.073903} {\bibfield  {journal} {\bibinfo  {journal}
  {Phys. Rev. Lett.}\ }\textbf {\bibinfo {volume} {111}},\ \bibinfo {pages}
  {073903} (\bibinfo {year} {2013})}\BibitemShut {NoStop}%
\bibitem [{\citenamefont {Buddhiraju}\ \emph {et~al.}(2020)\citenamefont
  {Buddhiraju}, \citenamefont {Song}, \citenamefont {Papadakis},\ and\
  \citenamefont {Fan}}]{buddhiraju_nonreciprocal_2020}%
  \BibitemOpen
  \bibfield  {author} {\bibinfo {author} {\bibfnamefont {S.}~\bibnamefont
  {Buddhiraju}}, \bibinfo {author} {\bibfnamefont {A.}~\bibnamefont {Song}},
  \bibinfo {author} {\bibfnamefont {G.~T.}\ \bibnamefont {Papadakis}}, \ and\
  \bibinfo {author} {\bibfnamefont {S.}~\bibnamefont {Fan}},\ }\href {\doibase
  10.1103/PhysRevLett.124.257403} {\bibfield  {journal} {\bibinfo  {journal}
  {Phys. Rev. Lett.}\ }\textbf {\bibinfo {volume} {124}},\ \bibinfo {pages}
  {257403} (\bibinfo {year} {2020})}\BibitemShut {NoStop}%
\bibitem [{\citenamefont {Bender}\ \emph {et~al.}(1999)\citenamefont {Bender},
  \citenamefont {Boettcher},\ and\ \citenamefont
  {Meisinger}}]{bender_PTsymmetric_1999}%
  \BibitemOpen
  \bibfield  {author} {\bibinfo {author} {\bibfnamefont {C.~M.}\ \bibnamefont
  {Bender}}, \bibinfo {author} {\bibfnamefont {S.}~\bibnamefont {Boettcher}}, \
  and\ \bibinfo {author} {\bibfnamefont {P.~N.}\ \bibnamefont {Meisinger}},\
  }\href {\doibase 10.1063/1.532860} {\bibfield  {journal} {\bibinfo  {journal}
  {Journal of Mathematical Physics}\ }\textbf {\bibinfo {volume} {40}},\
  \bibinfo {pages} {2201} (\bibinfo {year} {1999})}\BibitemShut {NoStop}%
\bibitem [{\citenamefont {Bender}(2007)}]{bender_making_2007}%
  \BibitemOpen
  \bibfield  {author} {\bibinfo {author} {\bibfnamefont {C.~M.}\ \bibnamefont
  {Bender}},\ }\href {\doibase 10.1088/0034-4885/70/6/R03} {\bibfield
  {journal} {\bibinfo  {journal} {Reports on Progress in Physics}\ }\textbf
  {\bibinfo {volume} {70}},\ \bibinfo {pages} {947} (\bibinfo {year}
  {2007})}\BibitemShut {NoStop}%
\bibitem [{\citenamefont {Kildal}(1990)}]{kildal_artificially_1990}%
  \BibitemOpen
  \bibfield  {author} {\bibinfo {author} {\bibfnamefont {P.-S.}\ \bibnamefont
  {Kildal}},\ }\href {\doibase 10.1109/8.59765} {\bibfield  {journal} {\bibinfo
   {journal} {IEEE Transactions on Antennas and Propagation}\ }\textbf
  {\bibinfo {volume} {38}},\ \bibinfo {pages} {1537} (\bibinfo {year}
  {1990})}\BibitemShut {NoStop}%
\bibitem [{\citenamefont {Kildal}\ and\ \citenamefont
  {Kishk}(2003)}]{kildal_em_2003}%
  \BibitemOpen
  \bibfield  {author} {\bibinfo {author} {\bibfnamefont {P.-S.}\ \bibnamefont
  {Kildal}}\ and\ \bibinfo {author} {\bibfnamefont {A.}~\bibnamefont {Kishk}},\
  }\href@noop {} {\bibfield  {journal} {\bibinfo  {journal} {Applied
  Computational Electromagnetics Society Journal}\ }\textbf {\bibinfo {volume}
  {18}},\ \bibinfo {pages} {32} (\bibinfo {year} {2003})}\BibitemShut {NoStop}%
\bibitem [{\citenamefont {Bliokh}\ \emph {et~al.}(2014)\citenamefont {Bliokh},
  \citenamefont {Bekshaev},\ and\ \citenamefont
  {Nori}}]{bliokh_extraordinary_2014}%
  \BibitemOpen
  \bibfield  {author} {\bibinfo {author} {\bibfnamefont {K.~Y.}\ \bibnamefont
  {Bliokh}}, \bibinfo {author} {\bibfnamefont {A.~Y.}\ \bibnamefont
  {Bekshaev}}, \ and\ \bibinfo {author} {\bibfnamefont {F.}~\bibnamefont
  {Nori}},\ }\href {\doibase 10.1038/ncomms4300} {\bibfield  {journal}
  {\bibinfo  {journal} {Nat Commun}\ }\textbf {\bibinfo {volume} {5}},\
  \bibinfo {pages} {3300} (\bibinfo {year} {2014})}\BibitemShut {NoStop}%
\bibitem [{\citenamefont
  {Silveirinha}(2019{\natexlab{a}})}]{silveirinha_time-reversal_2019}%
  \BibitemOpen
  \bibfield  {author} {\bibinfo {author} {\bibfnamefont {M.~G.}\ \bibnamefont
  {Silveirinha}},\ }\href {\doibase 10.3390/sym11040486} {\bibfield  {journal}
  {\bibinfo  {journal} {Symmetry}\ }\textbf {\bibinfo {volume} {11}},\ \bibinfo
  {pages} {486} (\bibinfo {year} {2019}{\natexlab{a}})}\BibitemShut {NoStop}%
\bibitem [{\citenamefont {Fernandes}\ and\ \citenamefont
  {Silveirinha}(2022)}]{fernandes_role_2022}%
  \BibitemOpen
  \bibfield  {author} {\bibinfo {author} {\bibfnamefont {D.~E.}\ \bibnamefont
  {Fernandes}}\ and\ \bibinfo {author} {\bibfnamefont {M.~G.}\ \bibnamefont
  {Silveirinha}},\ }\href {\doibase 10.1103/PhysRevApplied.18.024002}
  {\bibfield  {journal} {\bibinfo  {journal} {Phys. Rev. Appl.}\ }\textbf
  {\bibinfo {volume} {18}},\ \bibinfo {pages} {024002} (\bibinfo {year}
  {2022})}\BibitemShut {NoStop}%
\bibitem [{\citenamefont {Silveirinha}(2014)}]{silveirinha_trapping_2014}%
  \BibitemOpen
  \bibfield  {author} {\bibinfo {author} {\bibfnamefont {M.~G.}\ \bibnamefont
  {Silveirinha}},\ }\href {\doibase 10.1103/PhysRevA.89.023813} {\bibfield
  {journal} {\bibinfo  {journal} {Phys. Rev. A}\ }\textbf {\bibinfo {volume}
  {89}},\ \bibinfo {pages} {023813} (\bibinfo {year} {2014})}\BibitemShut
  {NoStop}%
\bibitem [{\citenamefont {Hsu}\ \emph {et~al.}(2016)\citenamefont {Hsu},
  \citenamefont {Zhen}, \citenamefont {Stone}, \citenamefont {Joannopoulos},\
  and\ \citenamefont {Soljačić}}]{hsu_bound_2016}%
  \BibitemOpen
  \bibfield  {author} {\bibinfo {author} {\bibfnamefont {C.~W.}\ \bibnamefont
  {Hsu}}, \bibinfo {author} {\bibfnamefont {B.}~\bibnamefont {Zhen}}, \bibinfo
  {author} {\bibfnamefont {A.~D.}\ \bibnamefont {Stone}}, \bibinfo {author}
  {\bibfnamefont {J.~D.}\ \bibnamefont {Joannopoulos}}, \ and\ \bibinfo
  {author} {\bibfnamefont {M.}~\bibnamefont {Soljačić}},\ }\href {\doibase
  10.1038/natrevmats.2016.48} {\bibfield  {journal} {\bibinfo  {journal} {Nat
  Rev Mater}\ }\textbf {\bibinfo {volume} {1}},\ \bibinfo {pages} {1} (\bibinfo
  {year} {2016})}\BibitemShut {NoStop}%
\bibitem [{\citenamefont {Alexoudi}\ \emph {et~al.}(2020)\citenamefont
  {Alexoudi}, \citenamefont {Kanellos},\ and\ \citenamefont
  {Pleros}}]{alexoudi_optical_2020}%
  \BibitemOpen
  \bibfield  {author} {\bibinfo {author} {\bibfnamefont {T.}~\bibnamefont
  {Alexoudi}}, \bibinfo {author} {\bibfnamefont {G.~T.}\ \bibnamefont
  {Kanellos}}, \ and\ \bibinfo {author} {\bibfnamefont {N.}~\bibnamefont
  {Pleros}},\ }\href {\doibase 10.1038/s41377-020-0325-9} {\bibfield  {journal}
  {\bibinfo  {journal} {Light Sci Appl}\ }\textbf {\bibinfo {volume} {9}},\
  \bibinfo {pages} {91} (\bibinfo {year} {2020})}\BibitemShut {NoStop}%
\bibitem [{\citenamefont {Haus}(1983)}]{haus_waves_1983}%
  \BibitemOpen
  \bibfield  {author} {\bibinfo {author} {\bibfnamefont {H.~A.}\ \bibnamefont
  {Haus}},\ }\href@noop {} {\emph {\bibinfo {title} {Waves and {Fields} in
  {Optoelectronics}}}}\ (\bibinfo  {publisher} {Prentice Hall},\ \bibinfo
  {address} {Englewood Cliffs, NJ},\ \bibinfo {year} {1983})\BibitemShut
  {NoStop}%
\bibitem [{\citenamefont {Grigoriev}\ \emph {et~al.}(2013)\citenamefont
  {Grigoriev}, \citenamefont {Tahri}, \citenamefont {Varault}, \citenamefont
  {Rolly}, \citenamefont {Stout}, \citenamefont {Wenger},\ and\ \citenamefont
  {Bonod}}]{grigoriev_optimization_2013}%
  \BibitemOpen
  \bibfield  {author} {\bibinfo {author} {\bibfnamefont {V.}~\bibnamefont
  {Grigoriev}}, \bibinfo {author} {\bibfnamefont {A.}~\bibnamefont {Tahri}},
  \bibinfo {author} {\bibfnamefont {S.}~\bibnamefont {Varault}}, \bibinfo
  {author} {\bibfnamefont {B.}~\bibnamefont {Rolly}}, \bibinfo {author}
  {\bibfnamefont {B.}~\bibnamefont {Stout}}, \bibinfo {author} {\bibfnamefont
  {J.}~\bibnamefont {Wenger}}, \ and\ \bibinfo {author} {\bibfnamefont
  {N.}~\bibnamefont {Bonod}},\ }\href {\doibase 10.1103/PhysRevA.88.011803}
  {\bibfield  {journal} {\bibinfo  {journal} {Phys. Rev. A}\ }\textbf {\bibinfo
  {volume} {88}},\ \bibinfo {pages} {011803} (\bibinfo {year}
  {2013})}\BibitemShut {NoStop}%
\bibitem [{\citenamefont {Baranov}\ \emph {et~al.}(2017)\citenamefont
  {Baranov}, \citenamefont {Krasnok},\ and\ \citenamefont
  {Alù}}]{baranov_coherent_2017}%
  \BibitemOpen
  \bibfield  {author} {\bibinfo {author} {\bibfnamefont {D.~G.}\ \bibnamefont
  {Baranov}}, \bibinfo {author} {\bibfnamefont {A.}~\bibnamefont {Krasnok}}, \
  and\ \bibinfo {author} {\bibfnamefont {A.}~\bibnamefont {Alù}},\ }\href
  {\doibase 10.1364/OPTICA.4.001457} {\bibfield  {journal} {\bibinfo  {journal}
  {Optica, OPTICA}\ }\textbf {\bibinfo {volume} {4}},\ \bibinfo {pages} {1457}
  (\bibinfo {year} {2017})}\BibitemShut {NoStop}%
\bibitem [{\citenamefont {Monticone}\ and\ \citenamefont
  {Alù}(2014)}]{monticone_embedded_2014}%
  \BibitemOpen
  \bibfield  {author} {\bibinfo {author} {\bibfnamefont {F.}~\bibnamefont
  {Monticone}}\ and\ \bibinfo {author} {\bibfnamefont {A.}~\bibnamefont
  {Alù}},\ }\href {\doibase 10.1103/PhysRevLett.112.213903} {\bibfield
  {journal} {\bibinfo  {journal} {Phys. Rev. Lett.}\ }\textbf {\bibinfo
  {volume} {112}},\ \bibinfo {pages} {213903} (\bibinfo {year}
  {2014})}\BibitemShut {NoStop}%
\bibitem [{\citenamefont {Lanneb{\`e}re}\ and\ \citenamefont
  {Silveirinha}(2015)}]{lannebere_optical_2015}%
  \BibitemOpen
  \bibfield  {author} {\bibinfo {author} {\bibfnamefont {S.}~\bibnamefont
  {Lanneb{\`e}re}}\ and\ \bibinfo {author} {\bibfnamefont {M.~G.}\ \bibnamefont
  {Silveirinha}},\ }\href {\doibase 10.1038/ncomms9766} {\bibfield  {journal}
  {\bibinfo  {journal} {Nature Communications}\ }\textbf {\bibinfo {volume}
  {6}},\ \bibinfo {pages} {8766} (\bibinfo {year} {2015})}\BibitemShut
  {NoStop}%
\bibitem [{\citenamefont {Monticone}\ \emph {et~al.}(2019)\citenamefont
  {Monticone}, \citenamefont {Sounas}, \citenamefont {Krasnok},\ and\
  \citenamefont {Alù}}]{monticone_can_2019}%
  \BibitemOpen
  \bibfield  {author} {\bibinfo {author} {\bibfnamefont {F.}~\bibnamefont
  {Monticone}}, \bibinfo {author} {\bibfnamefont {D.}~\bibnamefont {Sounas}},
  \bibinfo {author} {\bibfnamefont {A.}~\bibnamefont {Krasnok}}, \ and\
  \bibinfo {author} {\bibfnamefont {A.}~\bibnamefont {Alù}},\ }\href {\doibase
  10.1021/acsphotonics.9b01104} {\bibfield  {journal} {\bibinfo  {journal} {ACS
  Photonics}\ }\textbf {\bibinfo {volume} {6}},\ \bibinfo {pages} {3108}
  (\bibinfo {year} {2019})}\BibitemShut {NoStop}%
\bibitem [{\citenamefont {Silva}\ \emph {et~al.}(2020)\citenamefont {Silva},
  \citenamefont {Morgado},\ and\ \citenamefont
  {Silveirinha}}]{silva_multiple_2020}%
  \BibitemOpen
  \bibfield  {author} {\bibinfo {author} {\bibfnamefont {S.~V.}\ \bibnamefont
  {Silva}}, \bibinfo {author} {\bibfnamefont {T.~A.}\ \bibnamefont {Morgado}},
  \ and\ \bibinfo {author} {\bibfnamefont {M.~G.}\ \bibnamefont
  {Silveirinha}},\ }\href {\doibase 10.1103/PhysRevB.101.041106} {\bibfield
  {journal} {\bibinfo  {journal} {Phys. Rev. B}\ }\textbf {\bibinfo {volume}
  {101}},\ \bibinfo {pages} {041106} (\bibinfo {year} {2020})}\BibitemShut
  {NoStop}%
\bibitem [{\citenamefont {Prudêncio}\ and\ \citenamefont
  {Silveirinha}(2021)}]{prudencio_monopole_2021}%
  \BibitemOpen
  \bibfield  {author} {\bibinfo {author} {\bibfnamefont {F.~R.}\ \bibnamefont
  {Prudêncio}}\ and\ \bibinfo {author} {\bibfnamefont {M.~G.}\ \bibnamefont
  {Silveirinha}},\ }\href {\doibase 10.1063/5.0077123} {\bibfield  {journal}
  {\bibinfo  {journal} {Applied Physics Letters}\ }\textbf {\bibinfo {volume}
  {119}},\ \bibinfo {pages} {261101} (\bibinfo {year} {2021})}\BibitemShut
  {NoStop}%
\bibitem [{\citenamefont {Mann}\ \emph {et~al.}(2019)\citenamefont {Mann},
  \citenamefont {Sounas},\ and\ \citenamefont
  {Al{\`u}}}]{mann_nonreciprocal_2019}%
  \BibitemOpen
  \bibfield  {author} {\bibinfo {author} {\bibfnamefont {S.~A.}\ \bibnamefont
  {Mann}}, \bibinfo {author} {\bibfnamefont {D.~L.}\ \bibnamefont {Sounas}}, \
  and\ \bibinfo {author} {\bibfnamefont {A.}~\bibnamefont {Al{\`u}}},\ }\href
  {\doibase 10.1364/OPTICA.6.000104} {\bibfield  {journal} {\bibinfo  {journal}
  {Optica, OPTICA}\ }\textbf {\bibinfo {volume} {6}},\ \bibinfo {pages} {104}
  (\bibinfo {year} {2019})}\BibitemShut {NoStop}%
\bibitem [{\citenamefont {Morgado}\ and\ \citenamefont
  {Silveirinha}(2016)}]{morgado_single_2016}%
  \BibitemOpen
  \bibfield  {author} {\bibinfo {author} {\bibfnamefont {T.~A.}\ \bibnamefont
  {Morgado}}\ and\ \bibinfo {author} {\bibfnamefont {M.~G.}\ \bibnamefont
  {Silveirinha}},\ }\href {\doibase 10.1088/1367-2630/18/10/103030} {\bibfield
  {journal} {\bibinfo  {journal} {New J. Phys.}\ }\textbf {\bibinfo {volume}
  {18}},\ \bibinfo {pages} {103030} (\bibinfo {year} {2016})}\BibitemShut
  {NoStop}%
\bibitem [{\citenamefont {Silveirinha}\ \emph {et~al.}(2018)\citenamefont
  {Silveirinha}, \citenamefont {Gangaraj}, \citenamefont {Hanson},\ and\
  \citenamefont {Antezza}}]{silveirinha_fluctuation_2018}%
  \BibitemOpen
  \bibfield  {author} {\bibinfo {author} {\bibfnamefont {M.~G.}\ \bibnamefont
  {Silveirinha}}, \bibinfo {author} {\bibfnamefont {S.~A.~H.}\ \bibnamefont
  {Gangaraj}}, \bibinfo {author} {\bibfnamefont {G.~W.}\ \bibnamefont
  {Hanson}}, \ and\ \bibinfo {author} {\bibfnamefont {M.}~\bibnamefont
  {Antezza}},\ }\href {\doibase 10.1103/PhysRevA.97.022509} {\bibfield
  {journal} {\bibinfo  {journal} {Phys. Rev. A}\ }\textbf {\bibinfo {volume}
  {97}},\ \bibinfo {pages} {022509} (\bibinfo {year} {2018})}\BibitemShut
  {NoStop}%
\bibitem [{\citenamefont {Latioui}\ and\ \citenamefont
  {Silveirinha}(2019)}]{latioui_lateral_2019}%
  \BibitemOpen
  \bibfield  {author} {\bibinfo {author} {\bibfnamefont {H.}~\bibnamefont
  {Latioui}}\ and\ \bibinfo {author} {\bibfnamefont {M.~G.}\ \bibnamefont
  {Silveirinha}},\ }\href {\doibase 10.1103/PhysRevA.100.053848} {\bibfield
  {journal} {\bibinfo  {journal} {Phys. Rev. A}\ }\textbf {\bibinfo {volume}
  {100}},\ \bibinfo {pages} {053848} (\bibinfo {year} {2019})}\BibitemShut
  {NoStop}%
\bibitem [{\citenamefont {Suh}\ \emph {et~al.}(2004)\citenamefont {Suh},
  \citenamefont {Wang},\ and\ \citenamefont {Fan}}]{suh_temporal_2004}%
  \BibitemOpen
  \bibfield  {author} {\bibinfo {author} {\bibfnamefont {W.}~\bibnamefont
  {Suh}}, \bibinfo {author} {\bibfnamefont {Z.}~\bibnamefont {Wang}}, \ and\
  \bibinfo {author} {\bibfnamefont {S.}~\bibnamefont {Fan}},\ }\href {\doibase
  10.1109/JQE.2004.834773} {\bibfield  {journal} {\bibinfo  {journal} {IEEE
  Journal of Quantum Electronics}\ }\textbf {\bibinfo {volume} {40}},\ \bibinfo
  {pages} {1511} (\bibinfo {year} {2004})}\BibitemShut {NoStop}%
\bibitem [{\citenamefont
  {Silveirinha}(2019{\natexlab{b}})}]{silveirinha_hidden_2019}%
  \BibitemOpen
  \bibfield  {author} {\bibinfo {author} {\bibfnamefont {M.~G.}\ \bibnamefont
  {Silveirinha}},\ }\href {\doibase 10.1364/OE.27.014328} {\bibfield  {journal}
  {\bibinfo  {journal} {Opt. Express, OE}\ }\textbf {\bibinfo {volume} {27}},\
  \bibinfo {pages} {14328} (\bibinfo {year} {2019}{\natexlab{b}})}\BibitemShut
  {NoStop}%
\end{thebibliography}%

\end{document}